\documentclass[11pt]{article}
\usepackage{graphicx}
\usepackage{amsmath}
\usepackage{amsfonts}
\usepackage{amssymb}
\usepackage{epsfig}
\usepackage{times}
\usepackage{cite}
\usepackage{calc}
\usepackage{version}
\usepackage[english]{babel}
\usepackage{epsf}
\usepackage{graphics}
\usepackage{ulem}
\usepackage[dvips]{color}

\newcommand{\nc}{N_{_c}}
\newcommand{\cf}{C_{_F}}
\newcommand{\tr}{T_{_R}}
\newcommand{\nf}{n_{f}}
\newcommand{\ns}{N_{s}}
\newcommand{\calp}{{\cal P}}
\newcommand{\lqcd}{\Lambda_{_{\rm QCD}}}

\definecolor{headcolor}{rgb}{0.05, 0.05, 0.5} 
\definecolor{footcolor}{rgb}{0.05, 0.05, 0.5} 
\definecolor{Cboxcolor}{rgb}{0.7, 0.9, 1.0}
\definecolor{CCboxcolor}{rgb}{0.5, 0.9,1.0}
\definecolor{Yboxcolor}{rgb}{1.0,1.0,0.8}
\definecolor{Tschwarz}{rgb}{0, 0, 0 }       
\definecolor{Trot}{rgb}{1.0, 0.0, 0.0}   
\definecolor{Tlrot}{rgb}{0.6, 0.2, 0.2}       
\definecolor{Tgruen}{rgb}{0.1, 0.7, 0.1}
\definecolor{Tblau}{rgb}{0.05, 0.05, 0.5}
\definecolor{Tlblau}{rgb}{0.05, 0.75, 1.0}
\definecolor{Tgelb}{rgb}{0.95, 0.95, 0.6}
\definecolor{Tgold}{rgb}{0.95, 0.9, 0.3}
\definecolor{Tweiss}{rgb}{1.0, 1.0, 1.0}

\begin{document}

\begin{titlepage}

\null

\vskip 1.5cm

\vskip 1.cm

{\bf\large\baselineskip 20pt
\begin{center}
\begin{Large}
In-medium jet shape from energy collimation in parton showers: Comparison with CMS PbPb data at 2.76 TeV
\end{Large}
\end{center}
}
\vskip 1cm

\begin{center}
Redamy P\'erez-Ramos 
\footnote{e-mail: redamy.r.perez-ramos@jyu.fi}\\
\smallskip
Department of Physics, P.O. Box 35, FI-40014 University of Jyv\"askyl\"a, Jyv\"askyl\"a, Finland\\%
\bigskip
Thorsten Renk
\footnote{e-mail: trenk@phy.duke.edu}\\
\smallskip
Department of Physics, P.O. Box 35, FI-40014 University of Jyv\"askyl\"a, Jyv\"askyl\"a, Finland,\\%
Helsinki Institute of Physics, P.O. Box 64, FI-00014 University of Helsinki, Helsinki, Finland
\end{center}

\baselineskip=15pt

\vskip 3.5cm


{\bf Abstract}: We present the medium-modified energy collimation 
in the leading-logarithmic approximation (LLA) and next-to-leading-logarithmic approximation (NLLA)
of QCD. As a consequence of more accurate 
kinematic considerations in the argument of the  
Dokshitzer-Gribov-Lipatov-Altarelli-Parisi (DGLAP) fragmentation 
functions (FFs) we find a new NLLA correction ${\cal O}(\alpha_s)$ 
which accounts for the scaling violation of DGLAP FFs at small $x$.
The jet shape is derived from the energy collimation within the same approximations 
and we also compare our calculations for the energy collimation with the  
event generators Pythia 6 and YaJEM for the first  time in this paper. 
The modification of jets by the medium in both cases is implemented by altering 
the infrared sector using the Borghini-Wiedemann model.
The energy collimation and jet shapes qualitatively describe a clear broadening of 
showers in the medium, which is further supported by YaJEM in the final comparison of the jet shape
with CMS PbPb data at center-of-mass energy 2.76 TeV. The comparison of the biased 
versus unbiased YaJEM jet shape with the CMS data shows a more accurate agreement 
for biased showers and illustrates the importance of an accurate simulation of the experimental
jet-finding strategy.
\end{titlepage}

\section{Introduction}
The phenomenon of jet quenching was first established experimentally through the observed
suppression of high-$p_T$ hadrons in nucleus-nucleus (A-A)
collisions at the Relativistic Heavy Ion Collider (RHIC) and the 
Large Hadron Collider (LHC) \cite{dEnterria:2009am,Gyulassy:1993hr,Baier:1996sk,Zakharov:1997uu}.
It was then confirmed by various other measurements where highly virtual partons 
produced in hard processes in a medium showed modifications to the subsequent evolution of a QCD shower,
in particular softening and broadening of the resulting hadron distribution which leads to 
a reduction in the yield of leading hadrons and jets 
\cite{Adcox:2001jp,Adler:2003ii,Adare:2008ad,Andare:2012qi,Chatrchyan:2013kwa}.

In the vacuum, the production of highly virtual partons following 
the hard inelastic scattering of two partons from the incoming protons ($2\to2+X$) is 
followed by the fragmentation into a spray of hadrons
which are observed in high-energy collider experiments. The evolution of successive
splittings $q(\bar q)\to q(\bar q)g$, $g\to gg$ and $g\to q\bar q$ ($q$, $\bar q$ and $g$ 
label quark, antiquark and gluon respectively)
inside the parton shower prior to hadronization is well established and can be described  
by the DGLAP evolution equations for fragmentation 
functions in the leading-logarithmic approximation (LLA) 
of QCD \cite{Dokshitzer:1977sg,Gribov:1972ri,Altarelli:1977zs}
or alternatively in terms of Monte Carlo (MC) formulations such as the PYSHOW algorithm
\cite{Bengtsson:1986hr,Norrbin:2000uu}. In A-A collisions, partons 
produced in the hard inelastic scattering 
of two partons from nuclei ($2\to2+X$) propagate through the hot/dense 
QCD media also produced in such collisions
and their branching pattern is changed
by interacting with the color charges of the deconfined 
quark gluon plasma (QGP) \cite{Baier:2000mf}. As a consequence, additional medium-induced soft gluon radiation   
is produced in A-A collisions, which leads for instance to the modification of
high-$p_T$ hadroproduction \cite{Salgado:2002cd,Gyulassy:2001nm,Wang:2002ri}. 
At RHIC, the main observables considered to probe this physics 
were the nuclear suppression factor
of single inclusive hadrons $R_{AA}$ 
\cite{Shimomura:2005en,Afanasiev:2009aa} and the suppression factor  
of hard back-to-back dihadron correlations $I_{AA}$ \cite{Magestro:2005vm,Adams:2006yt}.
More recently, through the analytical developments of Refs.
\cite{MehtarTani:2011tz,MehtarTani:2010ma,Blaizot:2013vha}, it was 
demonstrated that medium-induced soft gluon radiation is ruled by a condition of
antiangular ordering over successive emissions of such gluons, which oppositely
to the condition of angular ordering in the vacuum (for a review see Ref. \cite{Khoze:1996dn}), 
leads to the jet broadening. However, further efforts 
are required in order to implement the full description of color decoherence effects in 
Monte Carlo event generators.

In this paper, we aim at discussing
two different observables currently relevant mainly for LHC physics: 
the energy collimation and the jet shape 
of medium-modified showers. We use a QCD-inspired model introduced by 
Borghini and Wiedemann (BW) \cite{Borghini:2005em} for the modification of the
fragmentation functions (FFs) by the medium
where the medium evolution itself is described by a hydrodynamical evolution. 
In the BW model, the DGLAP splitting
functions are enhanced in the infrared sector in order to mimic the 
medium-induced soft gluon radiation. In practice, the $1/z$ dependence of the
QCD vacuum splitting functions corresponding to the parton branchings
$q(\bar q)\to q(\bar q)g$ and $g\to gg$ are altered by introducing
a parameter $N_{\rm s}=1+f_{\rm med}$ ($f_{\rm med}\geq0$) in the form 
$P_{a\to bc}(z)=N_{s}/z+{\cal O}(1)$, which is simple and mostly leads to
analytical results. In Ref. \cite{Nayak:2009iw}, 
which appeared long after \cite{Borghini:2005em}, Nayak derived
in-medium expressions for the quark and gluon DGLAP splitting functions 
in nonequilibrium (nonisotropic) QCD at leading order in $\alpha_s$ as 
a function of arbitrary non-equilibrium distribution functions 
$f_q(\vec{p})$ and $f_g(\vec{p})$, where $\vec{p}$ is the momentum of the hard parton. 
The modification to the splitting functions turns out to be quite similar to the 
one introduced in the BW model \cite{Borghini:2005em} by some prefactor depending on 
$f_q(\vec{p})$ and $f_g(\vec{p})$, which affects both the infrared and 
regular terms of the evolution kernels. Later on, in Refs.
\cite{Blaizot:2013vha,Mehtar-Tani:2014yea}, the splitting functions were modified by an 
overall factor which depends on the pertinent medium parameters, such as the medium
transport coefficient $\hat{q}$. Whether these modification prefactors 
are related to those found in \cite{Nayak:2009iw} may 
be an interesting issue of further investigation, but this is out of the scope of this paper. 
Moreover, the BW model in Ref. \cite{Borghini:2005em} and the 
calculations in Ref. \cite{Mehtar-Tani:2014yea} show that the production of soft hadrons as 
described by the FFs is enhanced at small energy fraction $x$ of the 
outgoing hadrons. The same result was also found in Ref. \cite{Albino:2009hu}, where a full 
resummation from large to small $x$ was performed in the same frame of the vacuum Albino-Khiehl-Kramer
parametrization of FFs (for a detailed review see Ref. \cite{Albino:2008gy}), 
which further motivates the use of the simple
BW prescription in the following.

We start with the quantification of 
the jet energy collimation and a study of the jet broadening in gluon and quark jets in LLA.
The computation of the energy collimation 
was first performed analytically in the vacuum \cite{Dokshitzer:1991ej} 
and subsequently modeled in the medium \cite{PerezRamos:2012ci} by means of the inclusive 
spectrum of partons provided by the medium-modified solution of DGLAP FFs at 
large $x\sim1$. For this purpose, a high-energy jet of
half opening angle $\Theta_0$, energy $E$ and virtuality $Q=E\Theta_0$ 
produced in a nucleus-nucleus collision was considered, followed by the production of
one concentric sub-jet of opening angle $\Theta$ and transverse momentum $k_\perp=xE\Theta$ 
where the bulk $xE\sim E$ of the jet energy is contained \cite{Dokshitzer:1991ej,PerezRamos:2012ci}. 
By definition, the smaller the angle $\Theta$ where the jet energy 
is concentrated, the higher the jet energy collimation \cite{Dokshitzer:1991ej}. 
The half opening angle of the jet $\Theta_0$ should be fixed according to the jet 
definition used by the experiment. The energy collimation can be then determined by maximizing the distribution
of partons $D(x,E\Theta_0,E\Theta)$, which dominates the hard fragmentation ($x\sim1$) 
of the jet into a sub-jet, as discussed in Ref. \cite{PerezRamos:2012ci}. 

In this paper, we will provide a more accurate 
description of the energy collimation, which accounts for the $x$ dependence in the 
third argument of the FFs $D(x,\ln E\Theta_0,\ln xE\Theta)$. 
The account of the shift in $\ln x$ leads to a small next-to-leading logarithmic approximation (NLLA) 
correction of order ${\cal O}(\alpha_s)$ which decreases the energy collimation at intermediate values 
of $x$ as a consequence of the scaling violation of DGLAP FFs \cite{Nason:1993xx}. 
Our first aim is indeed to make a comparison for the jet energy collimation 
between the LLA, NLLA with Pythia 6 \cite{Bengtsson:1986hr,Norrbin:2000uu} 
and YaJEM \cite{Renk:2008pp} in the 
medium. Such studies have been done for the vacuum case \cite{Seymour:1997kj}, 
but it is far from being evident that there are no additional 
differences in the medium and this issue should be further studied. 

The integrated jet shape $\Psi(\Theta;\Theta_0)$ provides indeed an analogous 
measurement of how widely the transverse energy of the jet is spread. 
This observable was first studied in the vacuum in \cite{Seymour:1997kj} and generalized 
to the medium throughout the calculations presented in \cite{Salgado:2003rv} and 
\cite{Vitev:2008rz} in the framework of the cone and $k_t$ jet reconstruction algorithms.
By definition, $\Psi(\Theta;\Theta_0)$
determines the energy fraction [$x\equiv\Psi(\Theta;\Theta_0)$] of a jet of half opening angle 
$\Theta_0$ that falls into a sub-jet of half opening angle $\Theta$ for a fixed jet energy $E$. 
In our framework, we extract for the first time the angular dependence of the sub-jet energy fraction 
from the simple prescription provided by the LLA DGLAP energy collimation 
$x=f(E\Theta_0,E\Theta)$ at fixed energy $E$; no other computation 
for this observable is known in the context of LLA DGLAP 
evolution equations. Thus, the jet energy collimation is a LLA DGLAP observable; 
however it is not measured in a heavy-ions collisions context, but
jet shapes are. That is why, in the comparison with MC, we do the 
jet shape analysis with Pythia 6 and YaJEM 
using the FastJet package \cite{Cacciari:2011ma,Cacciari:2005hq} 
on the events and compare our results with CMS $pp$ and 
PbPb data for central collisions (0-10\%) at 2.76 TeV \cite{Chatrchyan:2013kwa}.
The BW prescription provides a simple test case for this, as it is analytically 
solvable and easily implemented in the YaJEM code, it will be shown to 
capture the main physics of additional soft gluon production and 
jet broadening through this observable.

Finally, for the purpose of a detailed comparison with data from the CMS experiment, 
the jet-shape analysis with YaJEM is performed for both 
PbPb and $pp$ collisions following the CMS analysis procedure closely. 
Jets are reconstructed with the anti-$k_T$ 
algorithm \cite{Cacciari:2011ma,Cacciari:2005hq} with a 
resolution parameter $R(\equiv\Theta_0)=0.3$. The clustering analysis is 
limited to charged particles with $e_i>1$ GeV inside the jet cone where 
$E_{\rm rec}\geq100$ GeV is required for a jet (i.e. $E_{\rm rec}$ stands for the recovered jet energy inside 
the cone) \cite{Chatrchyan:2013kwa}. The 
condition $e_i>1$ GeV removes the soft QCD medium background which may blur the jet 
fragmentation and jet-shape analysis. In order to illustrate the role of the bias caused by the jet-finding
procedure outlined above, 
we compare the biased jet shape (i.e. provided the CMS jet-finding conditions are fulfilled) with
an unbiased jet shape (which is a purely theoretical quantity) for both PbPb and $pp$ CMS data.

\section{Theoretical framework}
\subsection{Description of the process and kinematics}
\label{sec:theory}
\begin{figure}[ht]
\begin{center}
\epsfig{file=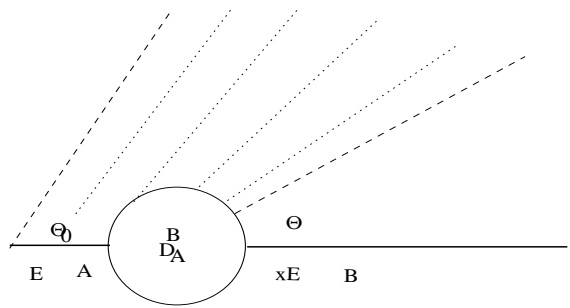, height=4.0truecm,width=7.5truecm}
\caption{\label{fig:energcollim} Fragmentation of a jet $A$ of half opening angle $\Theta_0$ into
a sub-jet $B$ of half opening angle $\Theta<\Theta_0$.}  
\end{center}
\end{figure}

In Fig.~\ref{fig:energcollim}, we consider the production of one gluon 
or quark ($A=g,q,\bar q$) jet of total energy $E$ and opening angle 
$\Theta_0$ which fragments into a sub-jet
$B$ of energy $xE$ and opening angle $\Theta$ ($\Theta\leq\Theta_0$), where
$x$ is the energy fraction of $A$ carried by $B$.

By definition, the virtualities of the jet $A$ and the sub-jet $B$ are 
$Q=E\Theta_0$ and $k_\perp=E_p\Theta$ ($E_p=xE$) respectively. The virtuality, also known as 
the hardness of the jet, 
determines the phase space for radiation  and hence sets the maximal transverse momentum
of a parton inside the jet: $k_\perp\leq Q$. A minimal cutoff parameter $Q_0$ can 
be introduced $k_\perp\geq Q_0$, such that the minimal angle reached by a parton
inside the cascade equals $\Theta_{\rm min}\geq Q_0/xE\Theta_0$. 
Experimentally, this physical picture
corresponds to the calorimetric measurement of the 
energy flux deposited within a given solid angle.
From the partonic point of view, the successive decays of 
partons in the cascade are ordered in $k_{\perp,i}$, 
or angles $\Theta_i$ due to the LLA kinematics for hard parton
decays ($x\sim1$) or due to the QCD coherence for soft 
parton decays ($x\ll1$) \cite{Dokshitzer:1991ej}. 
Hard parton decays determine the bulk of the jet energy and 
are ruled by the LLA kinematics, which leads to DGLAP evolution 
equations \cite{Dokshitzer:1978hw}, while
soft parton decays determine
the bulk of the jet multiplicity and are ruled by
QCD coherence, which leads instead to the modified-LLA (MLLA) evolution 
equations \cite{Dokshitzer:1991ej}.

The jet energy collimation is characterized by the large
energy fraction $x$ of the sub-jet where the bulk of the jet energy inside the
given cone $\Theta<\Theta_0\ll1$ is deposited. Hence, the probability for the energy fraction $x$
to be deposited in a cone of aperture $\Theta$ is related to the DGLAP inclusive spectrum
of partons through the formula \cite{Dokshitzer:1991wu},
\begin{equation}\label{eq:sumDAB}
D_A(x,E\Theta_0,xE\Theta)=\sum_{B=g,q}D_A^B(x,E\Theta_0,xE\Theta),
\end{equation}
where the nature of partons $B$ is not identified. In Eq.~(\ref{eq:sumDAB}), the FFs
$D_A^B(x,E\Theta_0,xE\Theta)$ determine the probability that a parton $A$ 
produced at large $p_T\sim E$ in a high-energy collision fragments into a hard sub-jet $B$
of transverse momentum $xE\Theta$, which we write in the third argument of the FF. 
Qualitatively, Eq.~(\ref{eq:sumDAB}) describes the evolution of the jet $A$ in the 
$k_\perp$ range $xE\Theta\leq k_\perp\leq E\Theta_0$ according to the
LLA $k_\perp$ ordering and hence, it determines the partonic skeleton of 
the sub-jet $B$ before the hadronization takes place. 

As compared to the FF for the inclusive spectrum of partons
where the third argument is set to $E\Theta$ for hard partons $x\sim1$: 
$D_A^B(x,E\Theta_0,E\Theta)$ \cite{PerezRamos:2012ci}, the formula 
(\ref{eq:sumDAB}) accounts for 
the energy fraction $x$ of the sub-jet $B$ in 
the FFs $D_A^B(x,E\Theta_0,xE\Theta)$. We cannot compute 
$D_A^B(x,E\Theta_0,xE\Theta)$ by using DGLAP evolution equations because of the $x$ dependence
included in the third argument, but we can instead expand it in powers of $``\ln x"$
through the exponential operator, 
\begin{equation}\label{eq:exptrick}
D_A^B(x,E\Theta_0,xE\Theta)=e^{\ln x(\partial/\partial\ln(E\Theta))} D_A^B(x,E\Theta_0,E\Theta)
\end{equation}
such that,
\begin{equation}
\label{eq:secondvers}
D_{A}^B(x,E\Theta_0,xE\Theta)= D_{A}^B(x,\Delta\xi)-
\ln x\frac{\alpha_s(E\Theta_0)}{2\pi}e^{4N_c\beta_0\Delta\xi}
\frac{\partial D_A^B}{\partial\Delta\xi}(x,\Delta\xi)+{\cal O }(\alpha_s^2),
\end{equation}
where,
\begin{equation}\label{eq:xivariables}
\xi(E\Theta)
=\frac1{4N_c\beta_0}\ln\left[\ln\left(\frac{E\Theta}{\lqcd}\right)^2\right],
\quad \Delta\xi=\xi(E\Theta_0)-\xi(E\Theta),
\quad \alpha_s(E\Theta_0)=\frac{2\pi}{4N_c\beta_0\ln\left(\frac{E\Theta_0}{\lqcd}\right)}.
\end{equation}
Hence, (\ref{eq:sumDAB}) can be rewritten in the form,
\begin{equation}\label{eq:sumDAB1}
 D_A(x,E\Theta_0,xE\Theta)\!=\!\!\sum_{B=g,q}\left[D_{A}^B(x,\Delta\xi)-
\ln x\frac{\alpha_s(E\Theta_0)}{2\pi}e^{4N_c\beta_0\Delta\xi}
\frac{\partial D_A^B}{\partial\Delta\xi}(x,\Delta\xi)\right]+{\cal O }(\alpha_s^2),
\end{equation}
where
$\alpha_s$ is the QCD coupling constant, $\beta_0$ is the first coefficient of the QCD $\beta$ function given 
by $\beta_0=\frac1{4N_c}\left(\frac{11}{3}N_c-\frac43T_Rn_f\right)$, with $N_c=3$, $T_R=\frac12$, $n_f=3$
and $\lqcd(=300$ MeV) is the mass scale of QCD. The new correction ${\cal O}(\alpha_s)$ is very small as $x\to1$ 
and can be much larger for $x\approx0.5$. As displayed in Fig.~\ref{fig:energcollim}, 
the ladder Feynman diagrams leading to DGLAP evolution equations for $D_{A}^B(x,\Delta\xi)$ should be iterated 
from the hardest virtuality $Q=E\Theta_0$ of the process to the lower 
sub-jet virtuality $k_\perp=E\Theta$ through the variable $\Delta\xi$ in (\ref{eq:xivariables}). 
Thus, $D_{A}^B(x,\Delta\xi)$
describes the distribution of partons $B$ with transverse momentum $k_\perp=E\Theta$ contained inside
the parton $A$, which fixes the initial scale of the hard process: $Q=E\Theta_0$; i.e. the virtuality.
Therefore, we can estimate $D_A(x,E\Theta_0,xE\Theta)$ 
with the solution of DGLAP evolution equations for the FFs $D_{A}^B(x,\Delta\xi)$, which 
appear on the rhs of Eq.~(\ref{eq:sumDAB1}).

\subsection{Medium-modified DGLAP evolution equations with the BW model}

The DGLAP evolution equations for the splitting $A[1]\to B[z]C[1-z]$ (where $z$ is the energy 
fraction of one parton in the splitting) in the 
$k_\perp$ range $E\Theta\leq k_\perp\leq E\Theta_0$ takes the simple form \cite{Dokshitzer:1978hw},
\begin{equation}\label{eq:dglapx}
\frac{d}{d\ln E\Theta}D(x,E\Theta_0,E\Theta)=\frac{\alpha_s(E\Theta)}{4\pi}\int_x^1\frac{dz}zP(z)
D\left(\frac{x}z,E\Theta_0,E\Theta\right),
\end{equation}
where $P(z)$ is the evolution ``Hamiltonian" given by
the regularized splitting functions \cite{Dokshitzer:1977sg,Gribov:1972ri,Altarelli:1977zs}. 
In order to account for the 
medium-induced soft gluon 
radiation in heavy-ion collisions, we make use of the QCD-inspired model 
proposed in Ref.~\cite{Borghini:2005em} which leads to
a simple solution of the evolution equations at $x\sim1$. In this model, the infrared 
parts of the splitting functions are arbitrarily enhanced by the factor $\ns=1+f_{\rm med}$,
where $f_{\rm med}\geq0$ accounts for medium-induced soft gluon radiation. 
The medium-modified splitting functions in $z$ space are written 
in the form \cite{Borghini:2005em},
\begin{subequations}
\begin{equation}\label{eq:splitG}
P_{gg}(z)=4\nc\left[\frac{\ns}{z}+\left[\frac{\ns}{1-z}\right]_+
+z(1-z)-2\right],\quad
P_{gq}(z)=2\ \tr[z^2+(1-z)^2],
\end{equation}
\begin{equation}\label{eq:splitQ}
P_{q g}(z)=2\ \cf\left(\frac{2\ns}{z}+z-2\right),\quad
P_{qq}(z)=2\ \cf\left(\left[\frac{2\ns}{1-z}\right]_+-1
-z\right),
\end{equation}
\end{subequations}
where the $[\dots]_+$ prescription is defined as $\int_0^1dx[F(x)]_+g(x)\equiv\int_0^1dxF(x)[g(x)-g(1)]$. 
The solutions of Eq.~(\ref{eq:dglapx}) can most conveniently be obtained in Mellin space 
${\cal D}(j,E\Theta_0,E\Theta)$ through the transformation
$$
{\cal D}(j,E\Theta_0,E\Theta)=\int_0^1dx\,x^{j-1}D(x,E\Theta_0,E\Theta),
$$
such that the convolution (\ref{eq:dglapx}) yields,
\begin{equation}\label{eq:dglapmellinconv}
\frac{d}{d\ln E\Theta}{\cal D}(j,E\Theta_0,E\Theta)={\cal P}(j){\cal D}(j,E\Theta_0,E\Theta),
\quad {\cal P}(j)=\int_0^1dz\,z^{j-1}P(z).
\end{equation}
The advantage of the Mellin transform can be clearly seen in Eq.~(\ref{eq:dglapmellinconv}). The 
convolution over $z$ in Eq.~(\ref{eq:dglapx}) reduces to the product of the Mellin-transformed
splitting functions ${\cal P}(j)$ and the FFs ${\cal D}(j,E\Theta_0,E\Theta)$.
Making use of the variables introduced in (\ref{eq:xivariables}),
(\ref{eq:dglapmellinconv}) can be more explicitly rewritten in the matrix form at leading order LO:
\begin{equation}\label{eq:dglap}
\frac{d}{d\xi}
\begin{pmatrix}
{\cal D}_{q_{\rm NS}}(j,\xi) \\
{\cal D}_{q_{\rm S}}(j,\xi) \\
{\cal D}_{g}(j,\xi)
\end{pmatrix} 
=\begin{pmatrix}
\calp_{qq}(j)&0&0 \\
0&\calp_{qq}(j)&\calp_{q g}(j)\\
0&\calp_{gq}(j)&\calp_{gg}(j) 
\end{pmatrix}
\begin{pmatrix}
{\cal D}_{q_{\rm NS}}(j,\xi) \\
{\cal D}_{q_{\rm S}}(j,\xi) \\
{\cal D}_{g}(j,\xi)
\end{pmatrix},
\end{equation}
where ${\cal D}_{q_{\rm NS}}$ and ${\cal D}_{q_{\rm S}}$ 
stand, respectively, for the flavor-nonsinglet 
and flavor-singlet quark distributions, and $\calp_{ik}(j)$ 
are the Mellin transforms of the LO splitting functions: 
\begin{subequations}\label{eq:splitmellin}
\begin{eqnarray}
\calp_{gg}(j)\!&\!=\!&\!-4 \nc\left[\ns\ \psi(j+1)+\ns\gamma_E-\frac{\ns-1}{j}-
\frac{\ns-1}{j-1}\right]\nonumber\\ 
&&+\frac{11 \nc}3-\frac{2\nf}3+\frac{8\nc(j^2+j+1)}{j(j^2-1)(j+2)}\label{eq:nuGbis},\\
\calp_{gq}(j)\!&\!=\!&\!\tr\ \frac{j^2+j+2}{j(j+1)(j+2)},\label{eq:splitjG} \\
\calp_{q g}(j)\!&\!=\!&\!2\ \cf\
\frac{(2 \ns-1)(j^2+j)+2}{j(j^2-1)}\label{eq:splitjQ},\\
\calp_{qq}(j)\!&\!=\!&\!-\cf\left[4\ns\ \psi(j+1)+4\ns\gamma_E-4\frac{\ns-1}j
-3-\frac2{j(j+1)}\right]\label{eq:nuFbis}.
\end{eqnarray}
\end{subequations}
This method 
allows for the diagonalization of the ``Hamiltonian" given by the set ${\cal P}(j)$ with 
respect to the ``evolution-time" variable $\xi\sim t=\ln(E\Theta)$. In some limits at 
large and small $x$, analytical solutions of the equations can be found through 
this method \cite{Dokshitzer:1978hw} but for numerical computation, solving 
the equations directly in $x$ space turns out to be more efficient than inverting 
the Mellin transform numerically.
Thus, at large energy fraction $x\sim1$, or equivalently large $j\gg1$ 
(which we are interested in), the expressions
for the Mellin representation of the splitting functions (\ref{eq:nuGbis})-(\ref{eq:nuFbis}) 
can be reduced to,
\begin{equation}\label{eq:approxdglapaa}
\calp_{qq}(j)\approx4C_FN_s\left(-\ln j+\frac3{4N_s}-\gamma_E\right),\quad
\calp_{gg}(j)\approx4N_cN_s\left(-\ln j+\frac{\beta_0}{N_s}-\gamma_E\right),
\end{equation}
where the asymptotic behavior of the digamma function $\psi(j+1)\approx\ln j$
is replaced at $j\gg1$ \cite{Dokshitzer:1978hw}.
The off-diagonal matrix elements vanish in this approximation: $\calp_{gq}(j)=\calp_{qg}(j)=0$.

Note that the $N_s\ln j$ dependence in Eq.~(\ref{eq:approxdglapaa}) arises from the $[N_s/(1-z)]_+$ terms of 
the DGLAP splitting functions, such that for hard partons $z\sim1$, the enhanced contribution of
the soft $1-z\sim0$ component $[1/(1-z)]_+$ produces the sub-jet broadening within this approximation.
Qualitatively, as a consequence of energy conservation, if soft
gluon radiation is enhanced in the region $\Theta\leq\Theta'\leq\Theta_0$ for a fixed jet energy $E$, the 
sub-jet energy ($B,\,xE$) should be smaller compared to its value in the vacuum and the energy 
collimation should then decrease, i.e. $\Theta$ should increase. 

Going back to $x$ space requires taking the inverse Mellin transform given by
\begin{equation}\label{eq:Ddelta}
D(x,\Delta\xi)=\frac{1}{2\pi i}\int_Cd j\, x^{-j}{\cal D}(j,\Delta\xi),
\end{equation} 
where the contour $C$ in the complex plane is parallel to the imaginary axis and lies to the right of all 
singularities. Since we are interested in the large $j\gg1$ ($x\sim1$) approximation, we insert 
Eq.~(\ref{eq:approxdglapaa}) into Eq.~(\ref{eq:dglap}). After integrating the result, 
we get the medium-modified distribution at large $x\sim1$,
\begin{equation}\label{eq:valDAA}
D_{A}^A(x,\Delta\xi)\simeq(1-x)^{-1+4C_AN_s\Delta\xi}\,\frac{\exp[4C_AN_s(\frac3{4N_s}-\gamma_E)\Delta\xi]}
{\Gamma(4C_AN_s\Delta\xi)}.
\end{equation}
where $\beta_0$ is replaced by $3/4$ ($n_f=3$) in Eq.~(\ref{eq:approxdglapaa}). The corresponding
result in the vacuum for $N_s=1$ is given in Ref.~\cite{Dokshitzer:1978hw}.

Within this approximation, the parton initiating the jet $A$ is identical to that initiating
the sub-jet $B=A$, $C_A=N_c$ if $A$ is a gluon and $C_A=C_F=\frac43$ if $A$ is a quark. Indeed, the 
FF $D_{A}^A(x,\Delta\xi)$ in Eq.~(\ref{eq:valDAA}) describes the splittings $g\to gg$ and $q\to qg$ and as 
constructed, it neglects the others: $g\to q\bar q$ and $q\to gq$. Therefore, the sum over $B$ in 
Eq.~(\ref{eq:sumDAB1}) disappears such that,
\begin{equation}
\label{eq:newcollimform}
D_{A}^A(x,E\Theta_0,xE\Theta)= D_{A}^A(x,\Delta\xi)-
\ln x\frac{\alpha_s(E\Theta_0)}{2\pi}e^{4N_c\beta_0\Delta\xi}\frac{\partial D_A^A}{\partial\Delta\xi}(x,\Delta\xi)
+{\cal O }(\alpha_s^2).
\end{equation}
In a more accurate solution of this problem which could only be achieved numerically, the whole sum of 
the parton branchings given by $D_q(x,E\Theta_0,xE\Theta)=D_q^q(x,E\Theta_0,xE\Theta)+D_q^g(x,E\Theta_0,xE\Theta)$ 
for a quark jet and $D_g(x,E\Theta_0,xE\Theta)=D_g^g(x,E\Theta_0,xE\Theta)+D_g^q(x,E\Theta_0,xE\Theta)$ 
for a gluon jet, with the full resummed contribution of the 
soft gluon/collinear logarithms arising from the $N_s/z$ dependence of the splitting functions 
in the FO approach of DGLAP FFs \cite{Albino:2004xa,Albino:2009hu}. 

\subsection{Jet energy collimation}

As discussed in Ref.~\cite{Dokshitzer:1978hw}, the distribution (\ref{eq:valDAA}) presents a certain 
maximum at some angle $\Theta$ where the bulk of the jet energy is concentrated. 
The reason for this can be understood as follows: for $\Delta\xi\to0$, or 
$\Theta\to\Theta_0$, almost all of the energy is contained inside the cone $\Theta_0$ 
[i.e. $D\to\delta(1-x)$] and the probability distribution $D_A^A$ for $x\ne1$ should 
decrease. For $\Theta$ decreasing $\Theta\gg \lqcd/E$ [notice that the $x$ dependence was 
reabsorbed on the pre-exponential term in Eq.~(\ref{eq:exptrick})], the emission outside the 
cone $\Theta$ grows and the fragmentation probability 
decreases. Then, taking the first derivative over $\ln\Theta$ in Eq.~(\ref{eq:newcollimform})
leads to the NLLA (not to be confused with the MLLA) equation for $\Theta$:
\begin{eqnarray}
&&\left[\ln(1-x)+\frac3{4N_s}-\gamma_E-\psi(4C_AN_s\Delta\xi)\right]
\left(1-4N_c\beta_0e^{b\Delta\xi}\ln x\frac{\alpha_s(E\Theta_0)}{2\pi}\right)=\cr
&&4C_AN_se^{4N_c\beta_0\Delta\xi}\ln x\frac{\alpha_s(E\Theta_0)}{2\pi}\left[\ln(1-x)+
\frac3{4N_s}-\gamma_E-\psi(4C_AN_s\Delta\xi)\right]^2\cr
&&-4C_AN_se^{4N_c\beta_0\Delta\xi}\ln x\frac{\alpha_s(E\Theta_0)}{2\pi}\psi^{(1)}(4C_AN_s\Delta\xi),
\label{eq:newcollimformbis}
\end{eqnarray}
which is the main theoretical result of this section for medium $N_s\ne0$ and also vacuum $N_s=0$. 
We invert Eq.~(\ref{eq:newcollimformbis}) numerically in order to get the NLLA 
jet energy collimation $\Theta(x,E)$. 
In Eq.~(\ref{eq:newcollimformbis}), $\psi(x)$ is the digamma function and 
$\psi^{(1)}(x)=\frac{d\psi(x)}{dx}$ is the polygamma 
function of the first order, which is new in this context. Note that this is one 
correction; a more complete set of
corrections of the same order can be also added if, for instance, one considers the
next-to-leading-order corrections \cite{Furmanski:1980cm} to the approached splitting functions 
(\ref{eq:approxdglapaa}) in a more cumbersome approach of this problem. However, this 
term goes beyond DGLAP and corresponds to the so-called scaling violation in DGLAP fragmentation 
functions \cite{Nason:1993xx}. In our framework, this correction slightly increases the
available phase space 
from the hardest ($B,\,x\sim1$) to slightly softer partons ($B,\,x\sim0.5$) 
and is therefore expected to decrease the energy collimation or increase $\Theta$ at intermediate $x$.
As expected for harder partons $\ln x\sim0$, the above 
equation (\ref{eq:newcollimformbis}) reduces to the simpler one \cite{PerezRamos:2012ci},
\begin{equation}\label{eq:digammaeq}
\psi(4C_AN_s\Delta\xi)=\ln(1-x)+\frac3{4N_s}-\gamma_E.
\end{equation}
Symbolically, the inversion of the NLLA (\ref{eq:newcollimformbis}) 
and LLA (\ref{eq:digammaeq}) can be written for quark ($A=q,\bar q$) and gluon ($A=g$) jets 
in the simple form,
\begin{equation}\label{eq:nscollim}
\frac{\Theta_A}{\Theta_0}=\left(\frac{E\Theta_0}{\lqcd}\right)^{-\gamma_A(x,N_s)}.
\end{equation}
Setting $\ln x\to0$ in Eq.~(\ref{eq:newcollimformbis}), the LLA expression for $\gamma_A(x,N_s)$ 
is simply written in the form \cite{PerezRamos:2012ci}
\begin{equation}\label{eq:llaslope}
\gamma_A(x,N_s)=1-\exp{\left[-\frac{N_c\beta_0}{C_AN_s}\psi^{-1}\left(\ln(1-x)
+\frac{3}{4N_s}-\gamma_E\right)\right]},
\end{equation}
where $\psi^{-1}$ is the inverse of the digamma function.
The exponent $\gamma_A(x,N_s)$ provides indeed the medium-modified slope of 
the energy collimation as a function of $N_s$ for a fixed value of the sub-jet energy fraction $x$
and can be obtained numerically from the NLLA equation (\ref{eq:newcollimformbis}). 
In Table~\ref{table:gammaA}, we display the values of the NLLA and LLA slopes for 
$x=0.5$ and $x=0.8$, which are in agreement with the 
LLA (NLLA) DGLAP large sub-jet energy fraction $x$ approximation where these 
predictions should be tested. As $x\to0$, the fixed-order (FO) approach of the LLA fails to 
provide any reliable result.
\begin{table}[htb]
\begin{center}
\begin{tabular}{ccccccc}
\hline\hline
NLLA, LLA & $x=0.5$ & $x=0.8$ & NLLA, LLA & $x=0.5$ & $x=0.8$ \\ \hline
$\gamma_g(x,1.4)$ & 0.37 & 0.26  & $\gamma_g(x,1)$ & 0.54 & 0.38  &\\
$\gamma_q(x,1.4)$ & 0.67 & 0.50 & $\gamma_q(x,1)$ & 0.83 & 0.65 &\\
\hline\hline
\end{tabular}
\caption{NLLA and LLA values of the slope $\gamma_A(x,N_s)$ of the energy collimation 
for $N_s=1.4$ (medium) and $N_s=1$ (vacuum).}
\label{table:gammaA}
\end{center}
\end{table}

The new equation (\ref{eq:newcollimformbis}) cannot be rewritten like Eq.~(\ref{eq:nscollim}) but it can 
be solved numerically. 
From Table~\ref{table:gammaA}, one may wonder why the NLLA (\ref{eq:newcollimformbis}) 
and LLA (\ref{eq:digammaeq}) slopes of the 
energy collimation are the same\footnote{Notice that Table~\ref{table:gammaA} displays 
indeed 16 values for the slopes, but such numbers are identical 
for the NLLA and LLA energy collimation.}. Indeed, the coupling constant does not depend 
on the jet energy only, but rather on the product 
$E\Theta\!\gg\!\Lambda_{\rm QCD}$ through the term 
$\ln xe^{4N_c\beta_0\Delta\xi}\alpha_s(E\Theta_0)\sim\ln x\alpha_s(E\Theta)$ 
in Eq.~(\ref{eq:newcollimformbis}). As the jet energy $E$ increases, 
the sub-jet cone $\Theta$ decreases and $\alpha_s(E\Theta)$ should remain
roughly constant. Therefore, the NLLA and LLA curves of the jet energy collimation should stay
approximately parallel to each other asymptotically. 

We can see in  both cases that the nuclear suppression parameter $N_s$ decreases the slope
of the energy collimation, which translates into increasing the rate of the jet 
broadening asymptotically. In both medium and vacuum $\gamma_q>\gamma_g$, which physically means 
that quark jets are more collimated than gluon jets.
The same trends should be confirmed in the forthcoming analysis
of the jet energy collimation with the event generator YaJEM.

\section{Comparison with YaJEM and QGP hydrodynamics}

In order to gauge the impact of the approximations made in deriving the results of the preceding section, we compare 
them with results for jet energy collimation obtained in a MC formulation of the in-medium jet evolution. Within such a model, the parton initiating a jet $A$ does not have to be identical to that initiating the sub-ject $B$ and hence the full set of splittings $g \rightarrow gg$ and $g \rightarrow q\overline q$ is available for a gluon jet. In addition, exact energy-momentum conservation at every splitting vertex is enforced. 

In vacuum, the PYSHOW algorithm \cite{Bengtsson:1986hr,Norrbin:2000uu} is a well-tested numerical implementation of QCD shower simulations. For comparison with our analytic results, we use the Borghini-Wiedemann prescription implemented within the in-medium shower code \cite{Renk:2009nz}.

\subsection{The in-medium shower generator YaJEM}
\label{subsec:yajemdescription}

YaJEM is based on the PYSHOW algorithm, to which it reduces in the limit of no medium effects. It simulates the evolution of a QCD shower as an iterated series of splittings of a parent into two daughter partons $a \rightarrow  bc$ where the energy of the daughters are obtained as $E_b = z E_a$ and $E_c = (1-z) E_a$ and the virtuality of parent and daughters is ordered as $Q_a \gg Q_b, Q_c$. The  decreasing hard virtuality scale of partons provides splitting by splitting the transverse phase space for radiation, and the perturbative QCD evolution terminates once the parton virtuality reaches a lower value $Q_0 = 1$ GeV, at which point the subsequent evolution is considered to be nonperturbative hadronization.

The probability distribution to split at given $z$ is given by the same QCD splitting kernels and their medium modification in the BW prescription, which we have used above, i.e. Eq.~(\ref{eq:splitG}) and  Eq.~(\ref{eq:splitQ}); however in the explicit kinematics of the MC shower the singularities for $z\rightarrow 0$ or $z \rightarrow 1$ are outside of accessible phase space and no $[\dots]_+$ regularization procedure is needed.

We will refer to the implementation of the BW prescription for in-medium showers in the following as YaJEM+BW (note that this corresponds to the FMED scenario described in Ref. \cite{Renk:2009nz}). This is distinct from the default version of the code YaJEM, YaJEM-DE, which is tested against multiple observables at both RHIC and LHC (see e.g. Refs.~\cite{Renk:2011aa,Renk:2012cb,Renk:2012hz}) and is based on an explicit exchange of energy and momentum between jet and medium rather than a modification of splitting probabilities.

For a straightforward benchmark comparison with analytic results, a value of $f_{\rm med}$ can be chosen, the parton shower can be computed and stopped at the partonic level or evolved using the Lund model to the hadronic level, and then clustered using the anti-$k_T$ algorithm and properties like collimation or jet shapes can then be extracted.

In a MC treatment of the shower evolution, using a constant value of $f_{\rm med}$ to characterize the medium is not needed and in fact not realistic once a comparison with data is desired. Following the procedure in 
Ref.~\cite{Renk:2009nz}, the value of $f_{\rm med}$ is determined event by event by embedding the hard process into a hydrodynamical medium \cite{Renk:2011gj} starting from a binary vertex which is at $(x_0,y_0)$ and following an eikonal trajectory $\zeta$ through the medium evaluating the line integral

\begin{equation}
\label{E-fmed}
f_{med} = K_f \int d \zeta [\epsilon(\zeta)]^{3/4} (\cosh \rho(\zeta) - \sinh \rho(\zeta) \cos\psi).
\end{equation}

where $\epsilon$ is the local energy density of the hydrodynamical medium, $\rho$ the local flow rapidity and $\psi$ the angle between the flow and the direction of  parton propagation.  Events are then generated for a large number of random $(x_0,y_0)$ sampled from the transverse overlap profile

\begin{equation}
\label{E-Profile}
P(x_0,y_0) = \frac{T_{A}({\bf r_0 + b/2}) T_A(\bf r_0 - b/2)}{T_{AA}({\bf b})},
\end{equation}

where $T_{A}$ is a nuclear thickness function $T_{A}({\bf r})=\int dz \rho_{A}({\bf r},z)$ obtained from the Woods-Saxon density $\rho_{A}({\bf r},z)$, and all observables are averaged over a sufficiently large number of events. This leaves a single dimensionful  parameter $K_f$ characterizing the strength of the coupling between parton and medium which is tuned to reproduce the measured nuclear suppression factor $R_{AA}$ in central 200 GeV AuAu collisions (see  Ref.~\cite{Renk:2009nz}). In practice, this procedure leads to an $\langle f_{\rm med} \rangle \approx 0.4$ which we will use in the analytical expressions when a comparison with data is intended.

\subsection{Medium-modified jet energy collimation}
\label{subsec:collimanalysis}
In this section we compare our NLLA (\ref{eq:newcollimformbis}) and 
LLA \cite{PerezRamos:2012ci} predictions 
for the energy collimation with YaJEM+BW. 
The analysis is carried out for gluon and quark
jets independently and including all particles in an event, i.e.
no detector effects are simulated in this section.

We generate thousands of gluon and quark 
dijets (i.e. back-to-back jets) 
for different fixed values of  
the center-of-mass energy $\sqrt{s}$ taken in the range 
$\sqrt{s}=100-1200$ GeV. By doing so, we fix the energy of the
leading initial parton to be $E_{\rm lp}=\sqrt{s}/2$ for each member in the dijet.
The values of $E_{\rm lp}$ are thus not selected as in the standard procedure
by sampling the initial energy (or $p_T$) distribution of partons provided 
by parton distribution functions (PDFs) \cite{Lai:2010nw}, 
nuclear parton distribution functions (nPDFs) \cite{Eskola:1998df} and the LO matrix 
elements of the partonic cross section as we do later when comparing with data.

Jets are then reconstructed by using the anti-$k_T$ 
algorithm \cite{Cacciari:2011ma,Cacciari:2005hq} 
inside the cone radii $R=1.0$, $0.3$ ($\Theta_0$), 
in agreement with the hard collinear approximation 
where the NLLA and LLA predictions should be tested. Reconstructed
jets can be sorted by energy ($E_{\rm jet,1}>E_{\rm jet,2}>\ldots$) 
event-by-event such that the most 
energetic one ($E_{\rm jet,1}$) can be randomly 
selected from its pair ($E_{\rm jet,2}$) for the analysis. 
We purposely use the default algorithm used by all LHC experiments.
Our motivation to do so from this subsection for the energy collimation is 
based on the fact that we then compare our Pyhtia 6 and YaJEM+BW results for the jet shapes with the CMS data 
in Sec.~\ref{section:CMSdatacomparison}, where jets are reconstructed with the anti-$k_t$ algorithm 
(see Ref.~\cite{Chatrchyan:2013kwa}). Besides, the anti-$k_T$ is the most robust jet reconstruction
algorithm for $pp$ and PbPb collisions at the LHC with respect to underlying events 
and pileup.

\begin{figure}[h]
\begin{center}
\epsfig{file=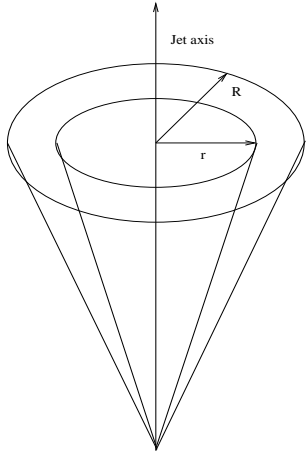, height=6.0truecm,width=4.0truecm}
\caption{\label{fig:jetshape} Jet ($\Theta_0=R$) and sub-jet ($\Theta=r$) 
cones sharing the bulk of the jet energy.}  
\end{center}
\end{figure}

The jet reconstruction radius coincides with 
the opening angle of the jet $\Theta_0=R$ and 
$\Theta=r$ with that of the sub-jet. 
The cone $R$ contains the reconstructed average energy flux ($E_{\rm rec}$) 
of the jet $A$ and $r$ the energy flux of the sub-jet $B$ ($xE_{\rm rec}$),
as illustrated in Fig.\ref{fig:jetshape}. 

The energy collimation can be then extracted from the angular distribution of the energy flux in $r< R$: 
$\frac{1}{N_{\rm jets}}\frac{d^2N}{dedr}$ for  
$r<R$ $(\Theta<\Theta_0)$. The recovered jet energy $E_{\rm rec}$ 
can be obtained by integrating $\frac{1}{N_{\rm jets}}\frac{d^2N}{dedr}$ 
over the whole range $0<r\leq R$ such that the fraction $x$ of the jet energy 
carried by the sub-jet $r<R$ can be written in the form:
\begin{equation}\label{eq:yajemcollim}
x=\frac1{E_{\rm rec}}\int_0^rdr'\int de\,e\frac{d^2N}{dedr'},
\quad E_{\rm rec}=\int_0^Rdr'\int de\,e\frac{d^2N}{dedr'}.
\end{equation}
By fixing the energy fraction $x$ in Eq.~(\ref{eq:yajemcollim})
to be large $x>0.5$, we can numerically compute the  
sub-jet radius $r$ where the bulk $xE_{\rm rec}$ of the reconstructed jet energy is 
contained. If the same procedure is repeated for different values of the center
of mass energy $\sqrt{s}$, the energy evolution of the collimation $r(E_{\rm rec})$ can be then displayed 
as a function of $E_{\rm rec}$ for a fixed sub-jet energy fraction $x$. 

The FastJet package \cite{Cacciari:2011ma,Cacciari:2005hq} provides 
the invariant mass $m_j=\sqrt{E_{j}^2-\vec{p}_j^{\,\,2}}$ of each jet 
independently from its pair if jets are sorted by the invariant mass 
($m_1>m_2>\ldots$), where $m_1$ is the invariant mass of the first jet and $m_2$ is 
the invariant mass of the second jet. The reconstructed virtuality of the jet inside small radii $R\ll1$ 
can be related to the invariant mass of the dijet $M_{jj}\gg m_j$ through
\begin{equation}\label{eq:invmass}
Q=\frac{M_{jj}}{2}R, 
\end{equation}
where $M_{jj}$ can be expressed in terms of each $m_j$,
\begin{equation}\label{eq:invmassdijet}
M_{jj}^2=m_1^2+m_2^2+2E_{1}E_{2}-2E_{1}E_{2}\sqrt{\left(1-\frac{m_1^2}{E_{1}^2}\right)
\left(1-\frac{m_2^2}{E_{2}^2}\right)}\cos\phi,
\end{equation}
which should be evaluated with $m_1\ne m_2$, $E_1\ne E_2$ and $\phi\approx\pi$ for clustered 
back-to-back jets. Indeed, the jet finder misses part of the initial jet energy $E_{\rm lp}$
and invariant mass $m_{\rm lp}$ such that, the new values inside $R$ are 
biased to smaller ones. Therefore, the reconstructed invariant mass of 
the dijet $M_{jj}$ should be estimated as given in Eq.~(\ref{eq:invmassdijet}) for the 
biased $E_i$ and $m_i$ obtained with FastJet.
The result in Eq.~(\ref{eq:invmass}) can be checked to be in good agreement with
the averaged one $Q=E_{\rm rec}R$ displayed in the Tables \ref{table:reccoeff}
and \ref{table:reccoeffq} in Secs. \ref{subsec:gluoncollim1} and
\ref{subsec:quarkcollim1}, respectively; the smaller the $R$ values, the more important the bias and
the better the agreement in the collinear limit, provided $R\geq \lqcd/E_{\rm rec}$. 
We rewrite Eq.~(\ref{eq:nscollim}) in the form,
\begin{equation}\label{eq:nscollim1}
r_A=R\left(\frac{Q}{\lqcd}\right)^{-\gamma_A(x,N_s)},\quad C_g=N_c=3,\quad C_q=C_F=\frac43,
\end{equation}
for the phenomenological treatment of the gluon ($C_g=N_c$) and quark ($C_q=C_F$) jets with $r_A\leq R$. 
We choose $\lqcd=300$ MeV, the same value as in the PYSHOW showering algorithm and do not use
the Lund model for the hadronization of partons into 
hadrons in this section \cite{Bengtsson:1986hr,Norrbin:2000uu}. 
The result for the energy collimation extracted from YaJEM+BW (\ref{eq:yajemcollim}) will be compared
with the medium-modified NLLA and LLA formulas (\ref{eq:newcollimformbis}). 
The medium modification parameter $f_{\rm med}$ is set to its mean value $\langle f\rangle_{\rm med}=0.4$,
obtained from averaging over the hydrodynamical model. 
The result of the numerical inversion of Eqs.~(\ref{eq:newcollimformbis}) 
and (\ref{eq:digammaeq}) will be displayed in the form given by Eq.~(\ref{eq:nscollim1}) in both cases.

\subsection{Medium-modified jet energy collimation in gluon jets}
 \label{subsec:gluoncollim1}
In Table~\ref{table:reccoeff}, we display gluon dijets at three different center-of-mass
energies $\sqrt{s}=150-500$ GeV, which are reconstructed by using the anti-$k_T$ algorithm 
\cite{Cacciari:2011ma,Cacciari:2005hq} for the radii
$R=1$ and $R=0.3$. As described above, $E_{\rm rec}$ is the recovered jet energy of the leading parton 
$E_{\rm rec}=\hat z\sqrt{s}/2$ inside the cone $R$ and $Q$ is the jet  virtuality. Note that the 
recovered energy of the leading parton equals the jet energy inside $R$ after reconstruction.
\begin{table}[htb]
\begin{center}
\begin{tabular}{cccccccc}
\hline\hline
$\sqrt{s}$ (GeV) & $R$
& $E_{\rm rec}$ (GeV) & $Q$ (GeV) & $R$ & $E_{\rm rec}$ (GeV) & $Q$ (GeV)\\ \hline
150 & 1.0 & 64.8 & 64.8 & $0.3$ & $46.3$ &  $13.8$ &\\
300 & 1.0 & 131.3 & 131.3 & $0.3$ & $98.0$ &  $29.4$ &\\
500 & 1.0 & 220.4 & 220.4 & $0.3$ & $168.9$ &  $50.7$ &\\
\hline\hline
\end{tabular}
\caption{Reconstructed jet energies inside the cone radii $R=1.0$ and $R=0.3$.}
\label{table:reccoeff}
\end{center}
\end{table}
\begin{figure}[h]
\begin{center}
\epsfig{file=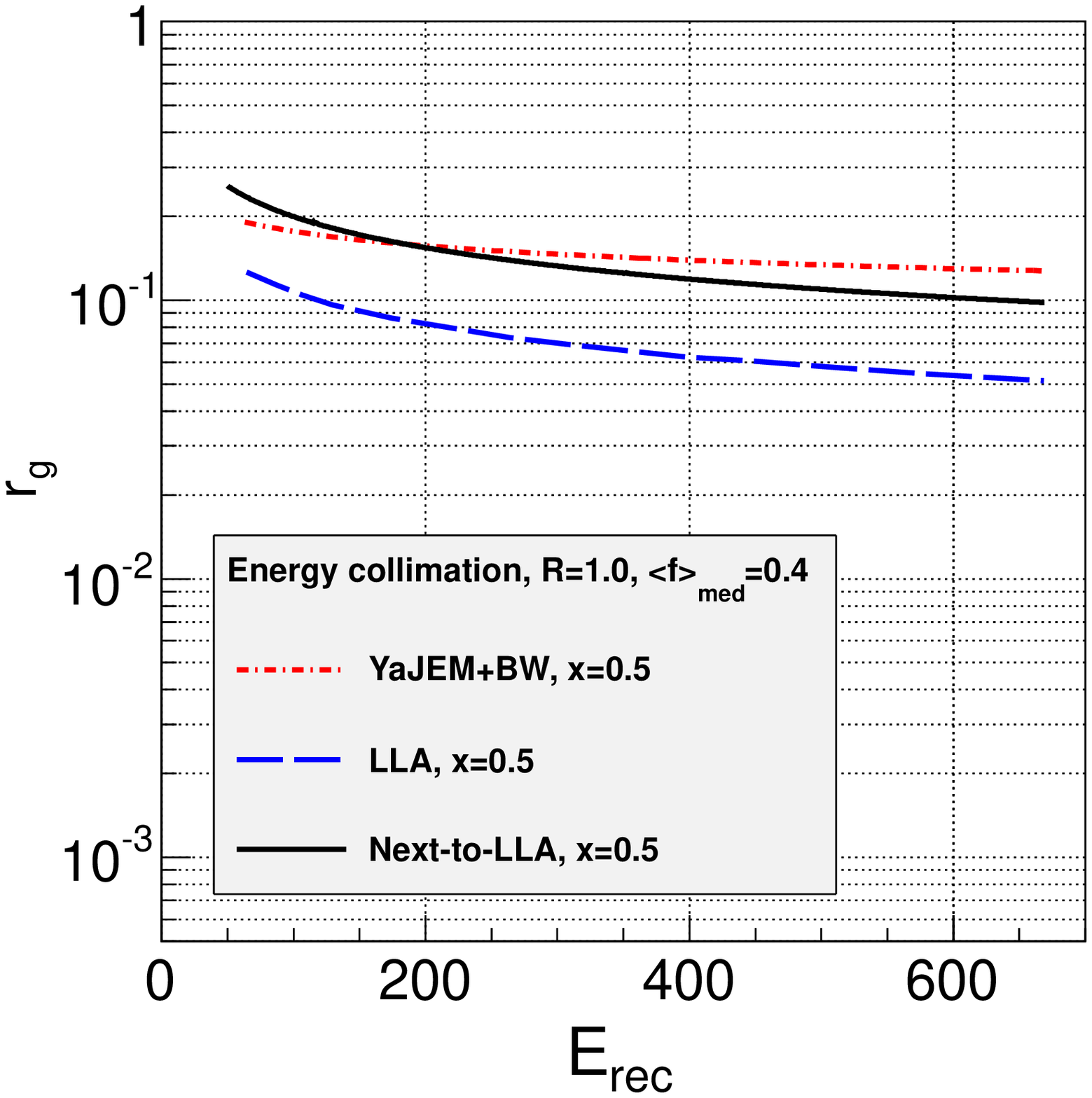, height=6.5truecm,width=7.8truecm}
\epsfig{file=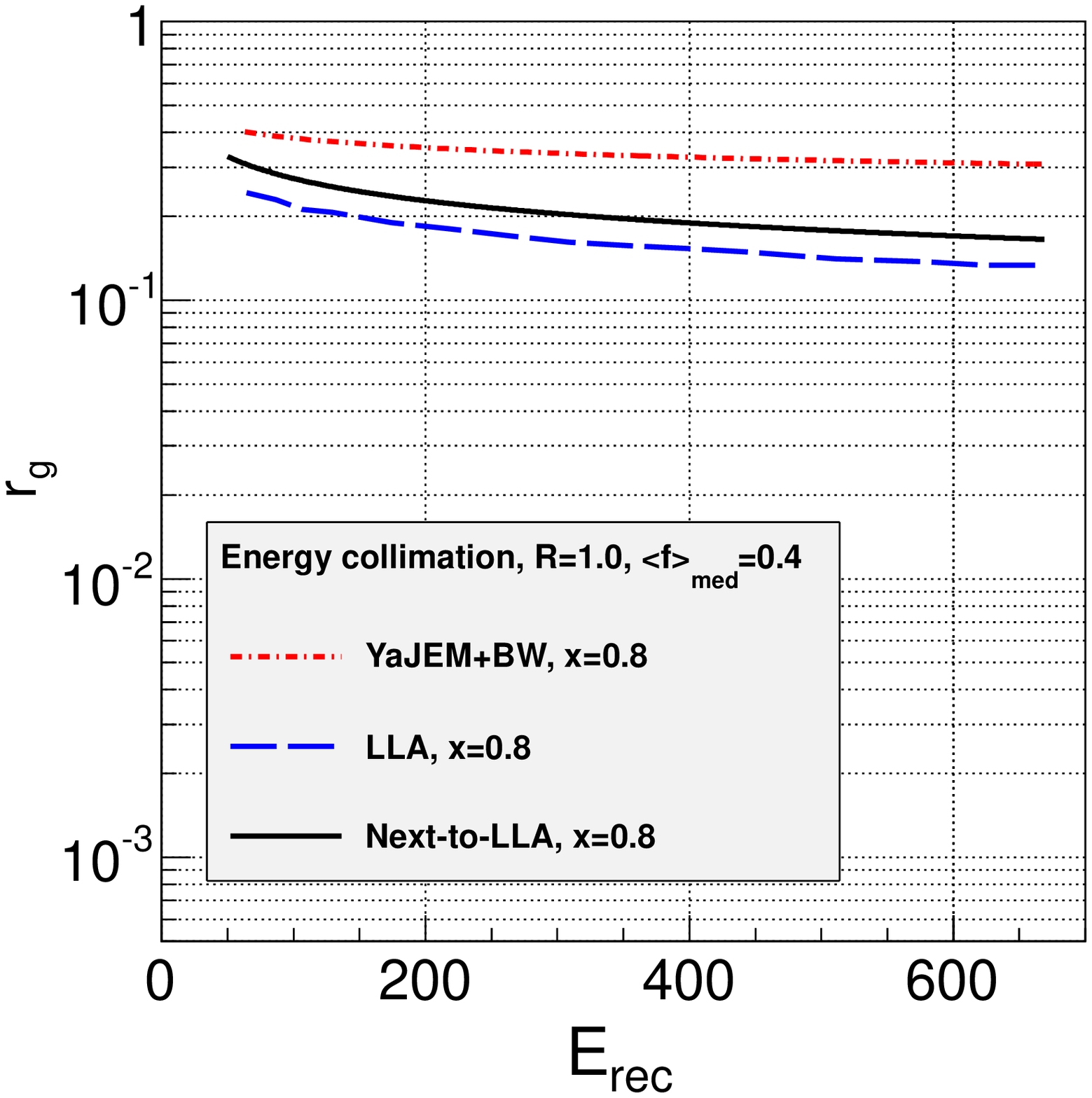, height=6.5truecm,width=7.8truecm}
\caption{\label{fig:collimgluon} Energy collimation inside a gluon jet for 
$x=0.5$ (left) and $x=0.8$ (right) with $R=1.0$.}  
\end{center}
\end{figure}
We can see that gluon jets carry the energy fractions $\hat{z}\sim4/5$ and $2/3$ of 
the leading parton for $R=1.0$ and $R=0.3$ respectively. Indeed, $\sqrt{s}/2$ is the energy of 
the leading parton spread in the whole hemisphere and $\sqrt{s}/2-E_{\rm rec}$ is that part of 
the jet energy that is lost outside $R$, which should not be confused with the jet energy 
inside the cone shell $r\leq r'\leq R$. We can see that the 
correlation between parton kinematics and reconstructed jet kinematics
gets increasingly blurred for small reconstruction radii. Thus,
the choice $R=0.3$ used by the CMS experiment provides a more severely 
biased jet energy.

In Fig.\ref{fig:collimgluon}, 
we display the energy collimation at $R=1.0$ in gluon 
jets,
\begin{equation}\label{eq:nscollimg1}
r_g=R\left(\frac{Q}{\lqcd}\right)^{-\gamma_g(x,N_s)}
\end{equation}
in the energy range $60\leq E_{\rm rec} (\text{GeV})\leq 600$ for RHIC and LHC phenomenology.
We choose the energy fractions $x=0.5$, $x=0.8$ and compare the NLLA prediction 
(\ref{eq:newcollimformbis}) with the LLA (\ref{eq:digammaeq}) and YaJEM+BW 
(\ref{eq:yajemcollim}) for $\langle f\rangle_{\rm med}=0.4$. The disagreement between 
the LLA prediction and YaJEM+BW is quite substantial and mainly due to the lack of other
perturbative contributions in this calculation, whereas the NLLA prediction improves the agreement. 
As expected for $x=0.5$, the ${\cal O}(\alpha_s)$ correction in the NLLA formula
(\ref{eq:newcollimformbis}) proves to be larger than for $x=0.8$. 
The shape of the energy collimation provided by the NLLA (\ref{eq:newcollimformbis}) and
the  LLA (\ref{eq:digammaeq}) are identical but steeper than the slope of 
the energy collimation provided by YaJEM+BW. Therefore, the NLLA and LLA
predictions overestimate the energy collimation compared to the YaJEM+BW prescription.

\begin{figure}[h]
\begin{center}
\epsfig{file=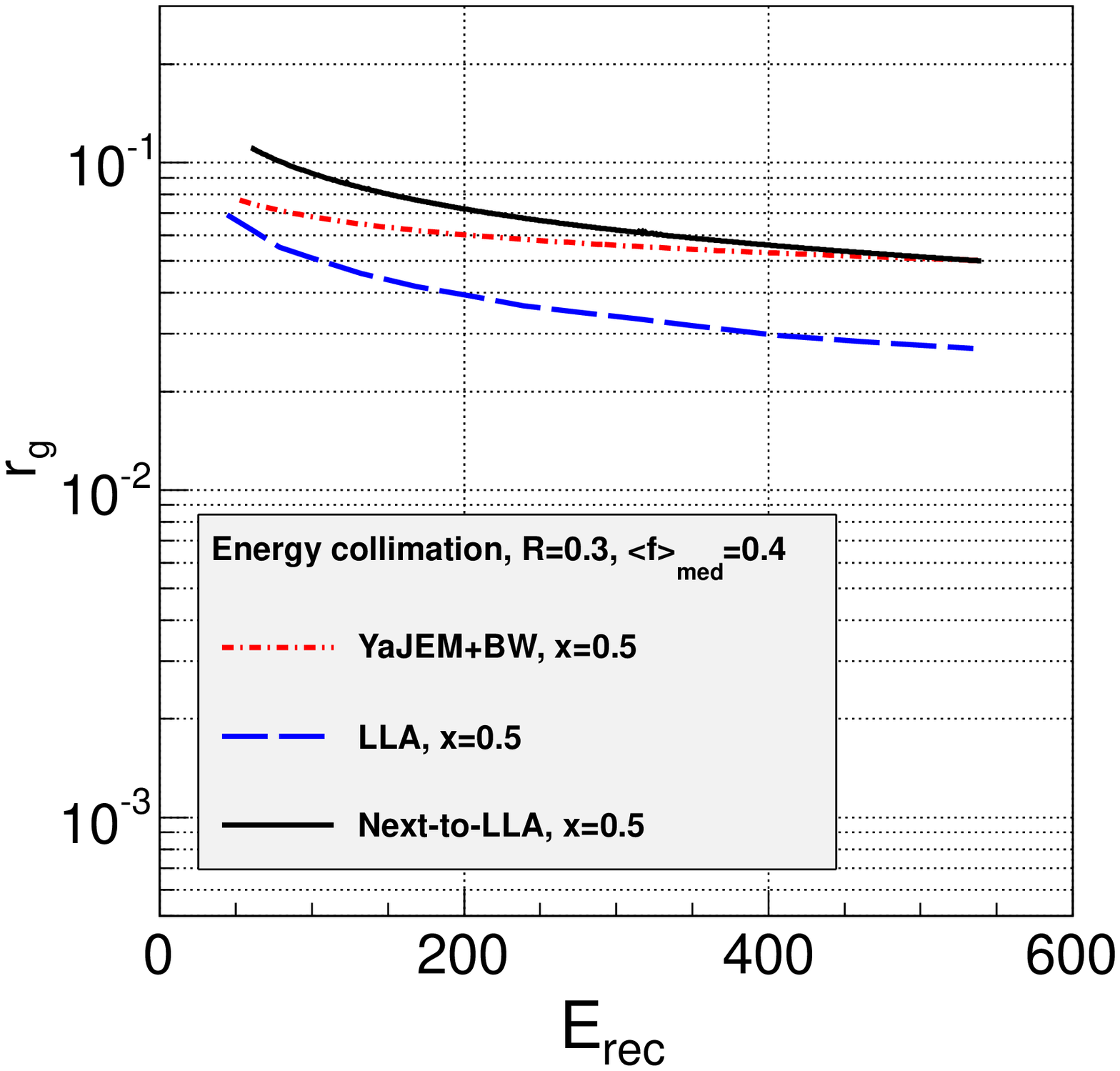, height=6.5truecm,width=7.8truecm}
\epsfig{file=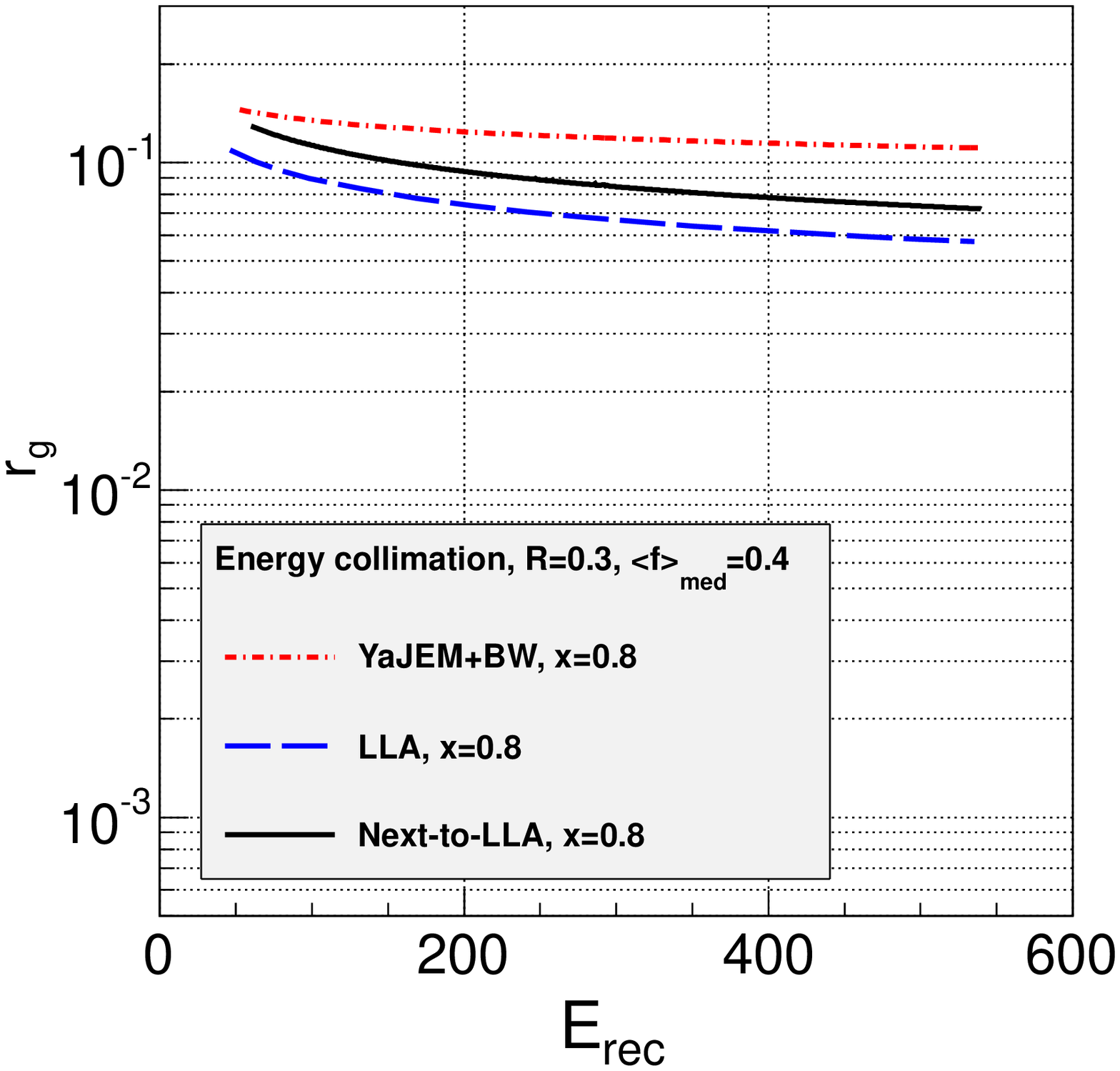, height=6.5truecm,width=7.8truecm}
\caption{\label{fig:collimgluonR03} Collimation of energy inside a gluon jet for 
$x=0.5$ (left) and $x=0.8$ (right) with $R=0.3$.}  
\end{center}
\end{figure}

Decreasing the jet radius to the value used by the CMS experiment at $2.76$ TeV PbPb 
collisions $R=0.3$ leads to a sizable hardening of the biased jet which may provide a 
better comparison between the NLLA, LLA and YaJEM+BW predictions 
for the jet energy collimation. The bias drives results to a generic outcome, so 
differences in the comparison must disappear as the bias gets stronger.
That is why, in Fig.\ref{fig:collimgluonR03} we display the same curves as in 
Fig.~\ref{fig:collimgluon} for $R=0.3$. We can see that the description provided by the 
NLLA (\ref{eq:yajemcollim}), LLA (\ref{eq:digammaeq}) and YaJEM+BW (\ref{eq:yajemcollim})
calculations are in better agreement with one another than the results displayed in 
Fig.~\ref{fig:collimgluon} for $R=1.0$. As expected, in the LLA, NLLA and YaJEM+BW 
computations, the energy collimation is stronger as the jet 
energy increases.

\begin{figure}[h]
\begin{center}
\epsfig{file=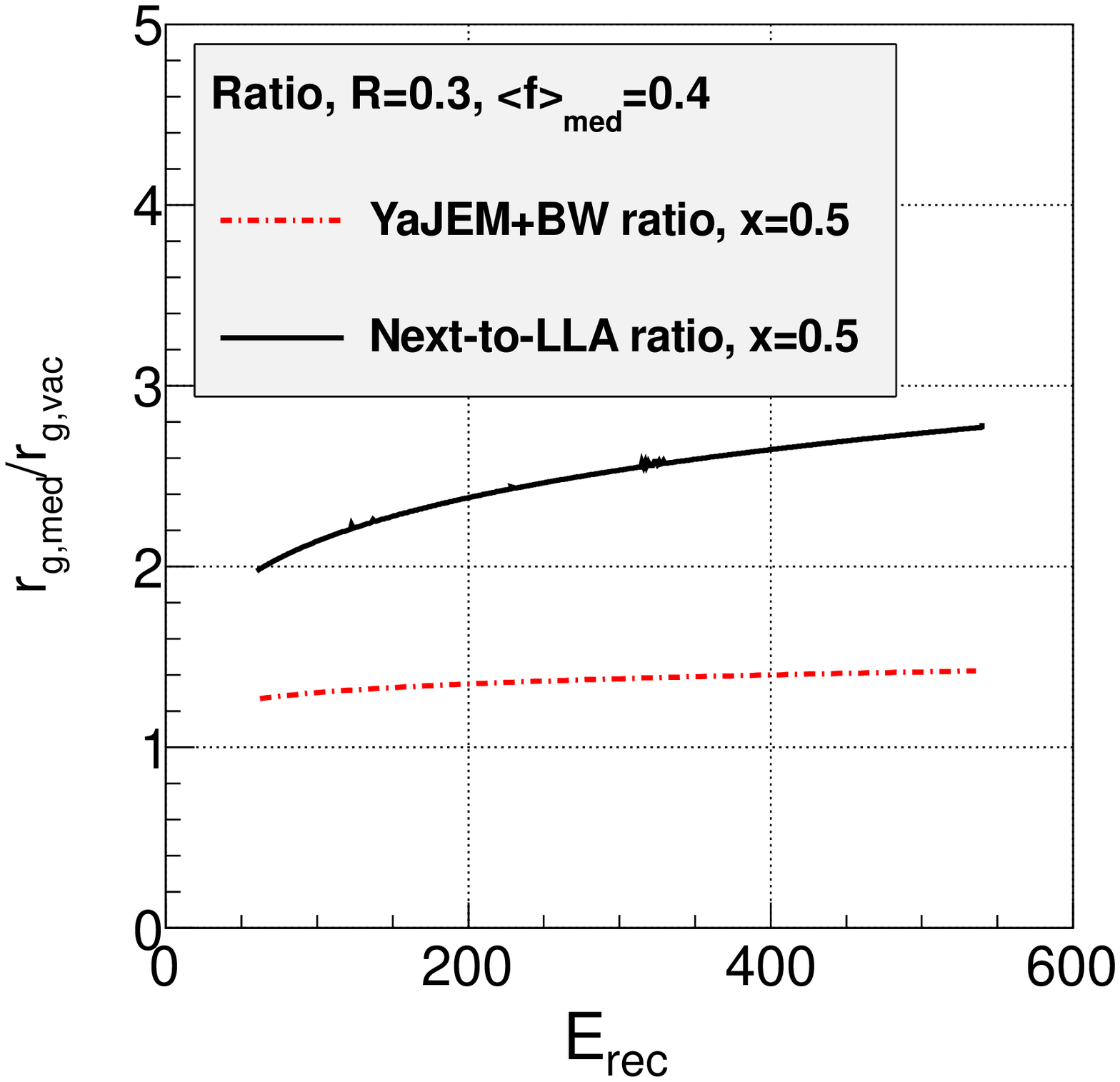, height=5.0truecm,width=5.2truecm}
\epsfig{file=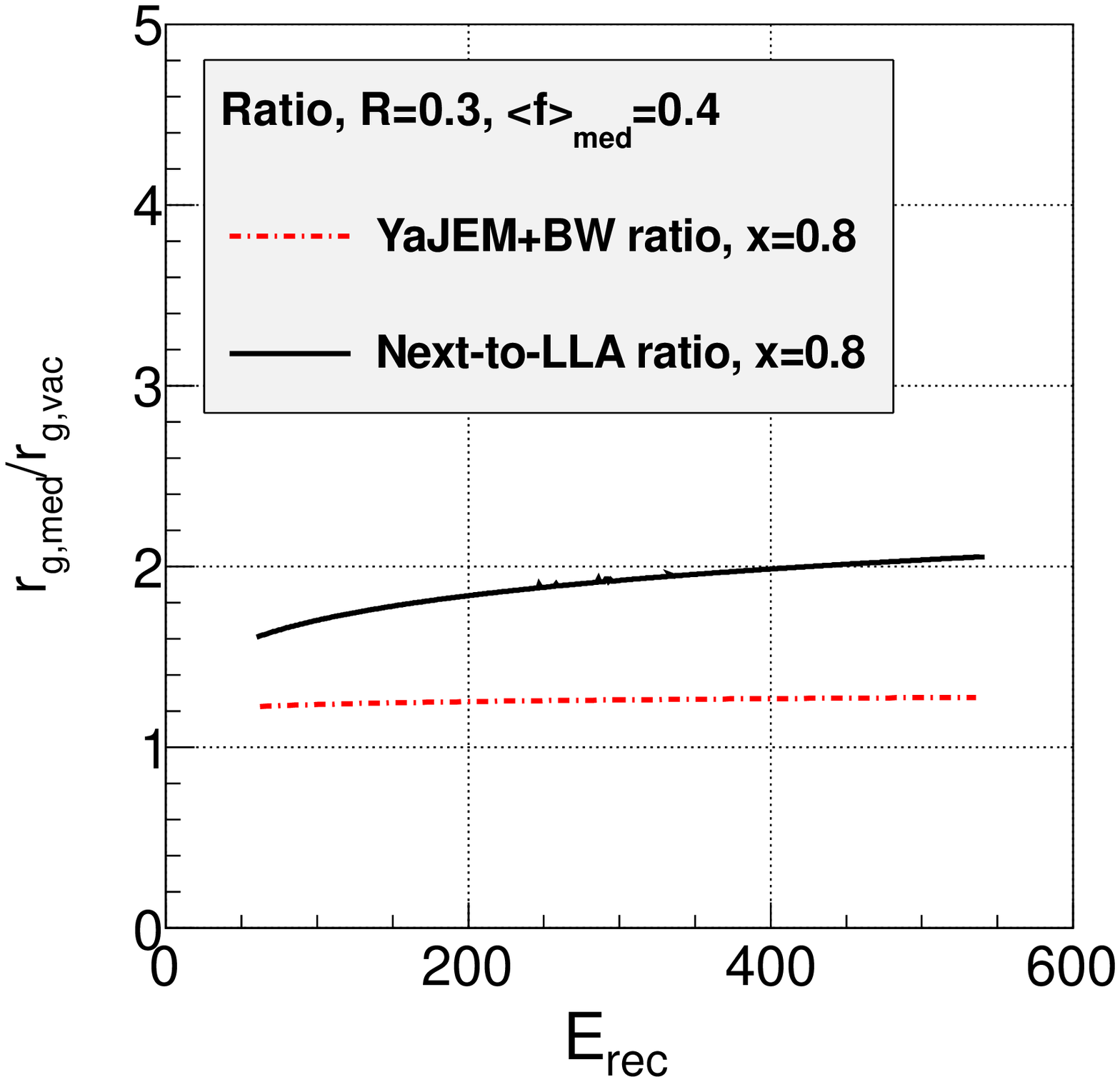, height=5.0truecm,width=5.2truecm}
\epsfig{file=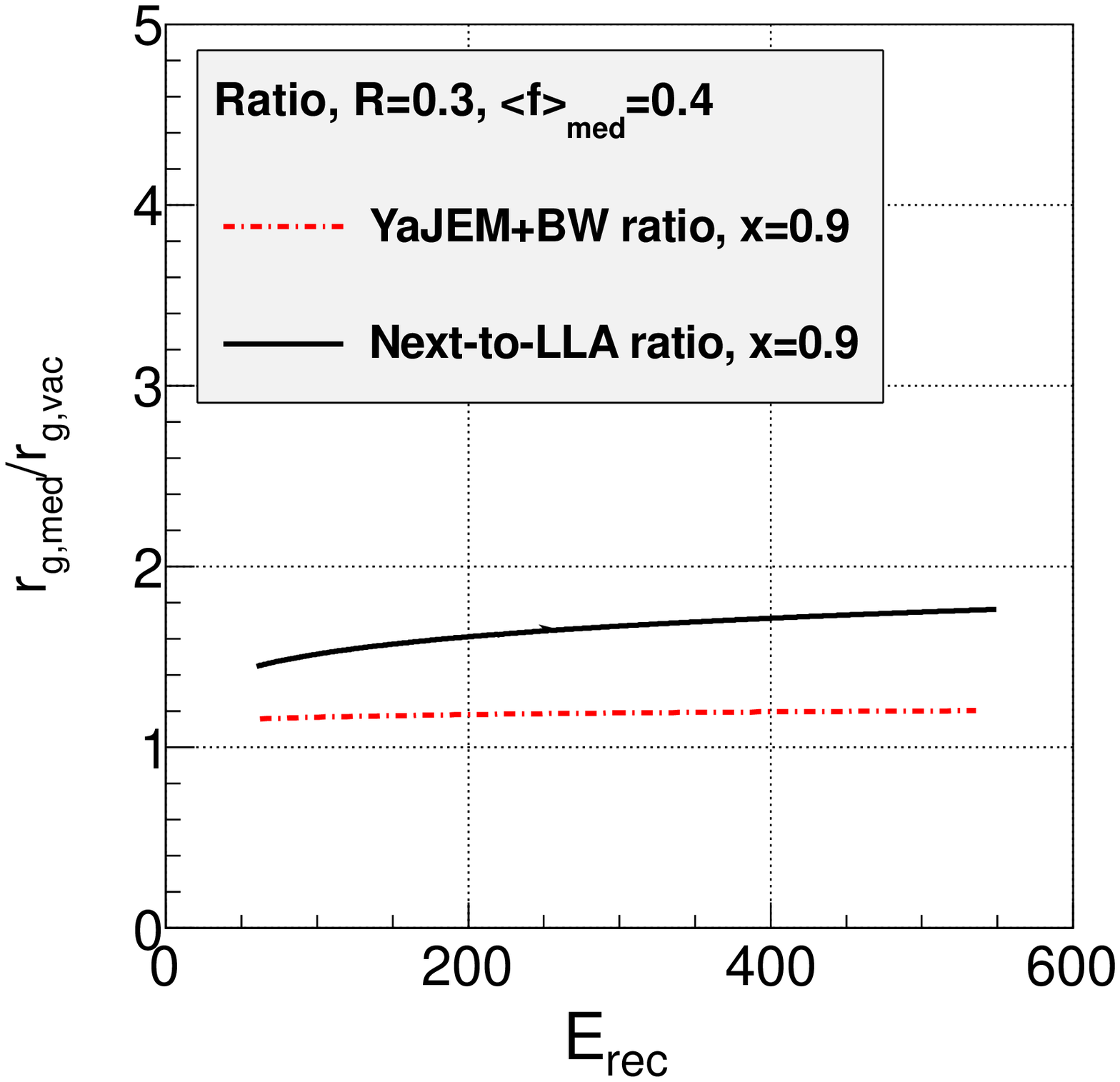, height=5.0truecm,width=5.2truecm}
\caption{\label{fig:ratiomedvsvacR03} Medium-modified and vacuum energy collimation ratios 
$r_{g,\rm med}/r_{g,\rm vac}$ for $x=0.5$ (left), 
$x=0.8$ (center) and $x=0.9$ (right) with $R=0.3$.}  
\end{center}
\end{figure}
As a consequence of jet quenching in high-energy heavy-ion collisions, medium-modified
showers are expected to broaden compared with vacuum showers. This effect can be 
quantified via the shower energy collimation by taking the ratios $r_{g,\rm med}/r_{g,\rm vac}$ with 
$\langle f\rangle_{\rm med}=0.4$ for the medium in 
the numerator and $\langle f\rangle_{\rm med}=0$ for vacuum in the denominator 
with the NLLA formula (\ref{eq:newcollimformbis}) and the YaJEM analysis (\ref{eq:yajemcollim}).
The ratios are displayed in Fig.~\ref{fig:ratiomedvsvacR03} for the energy fractions 
$x=0.5, 0.8, 0.9$ and the jet radius $R=0.3$. 
For $x=0.5$, the NLLA formula predicts a sub-jet broadening that is twice as large (i.e. smaller energy collimation)
than that in YaJEM+BW, while for $x=0.8-0.9$ the agreement is improved but still different by a factor of 
$\sim1.5-1.2$, reaching the best agreement for $x=0.9$. Thus, as the energy fraction $x$ increases, 
the jet broadening inside the smaller cone $r$ decreases. As expected, 
the NLLA correction seems to play a more important role as $x$ decreases. 
The latter can be observed in Fig.~\ref{fig:ratiomedvsvacR03} as 
one compares the shapes of the NLLA prediction with YaJEM+BW. 
The YaJEM+BW prediction tends to flatten while the NLLA formula increases,
making the vertical difference higher as the energy scale increases.

\subsection{Medium-modified jet energy collimation in quark jets}
\label{subsec:quarkcollim1} 
\begin{figure}[h]
\begin{center}
\epsfig{file=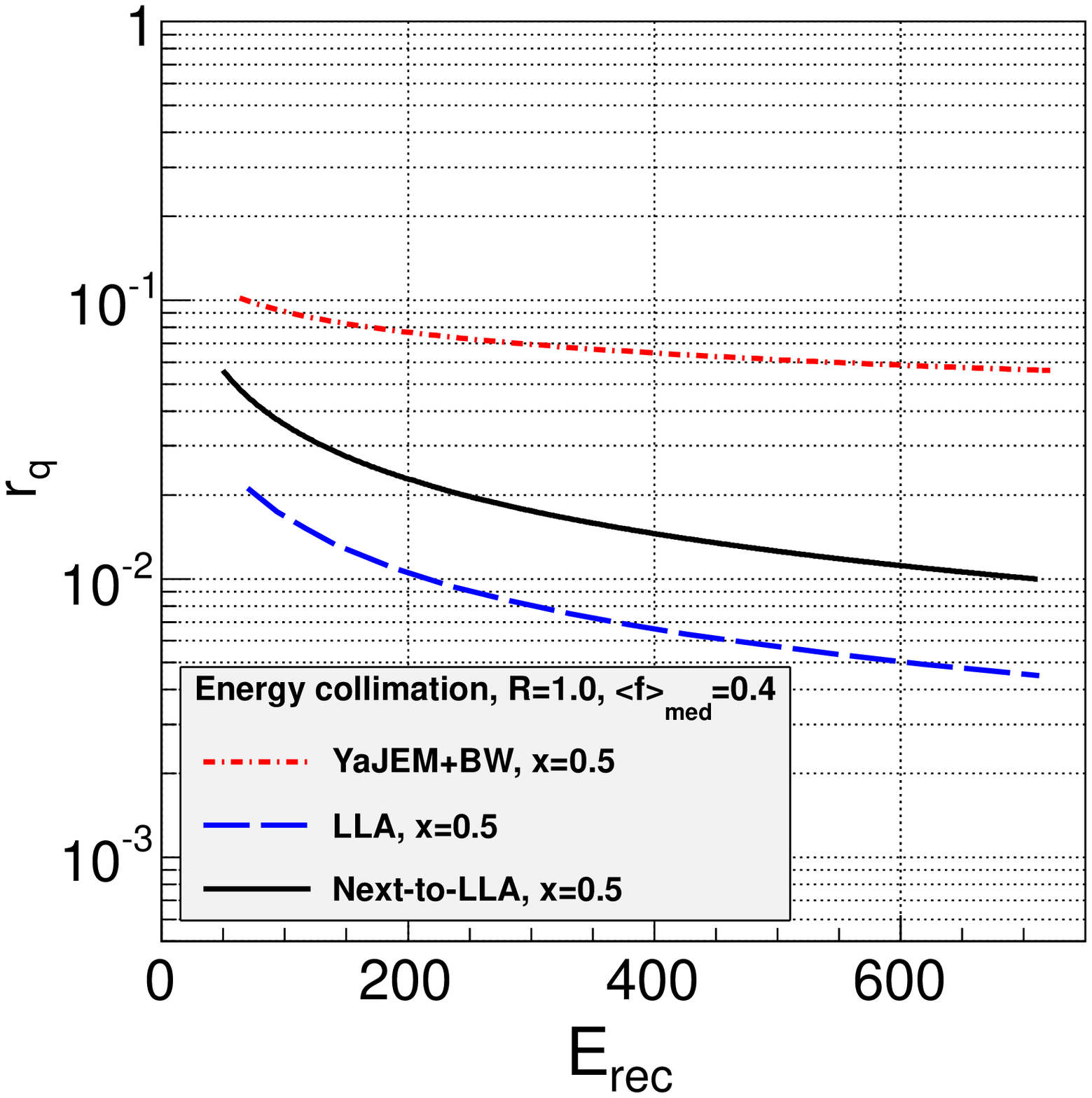, height=6.5truecm,width=7.8truecm}
\epsfig{file=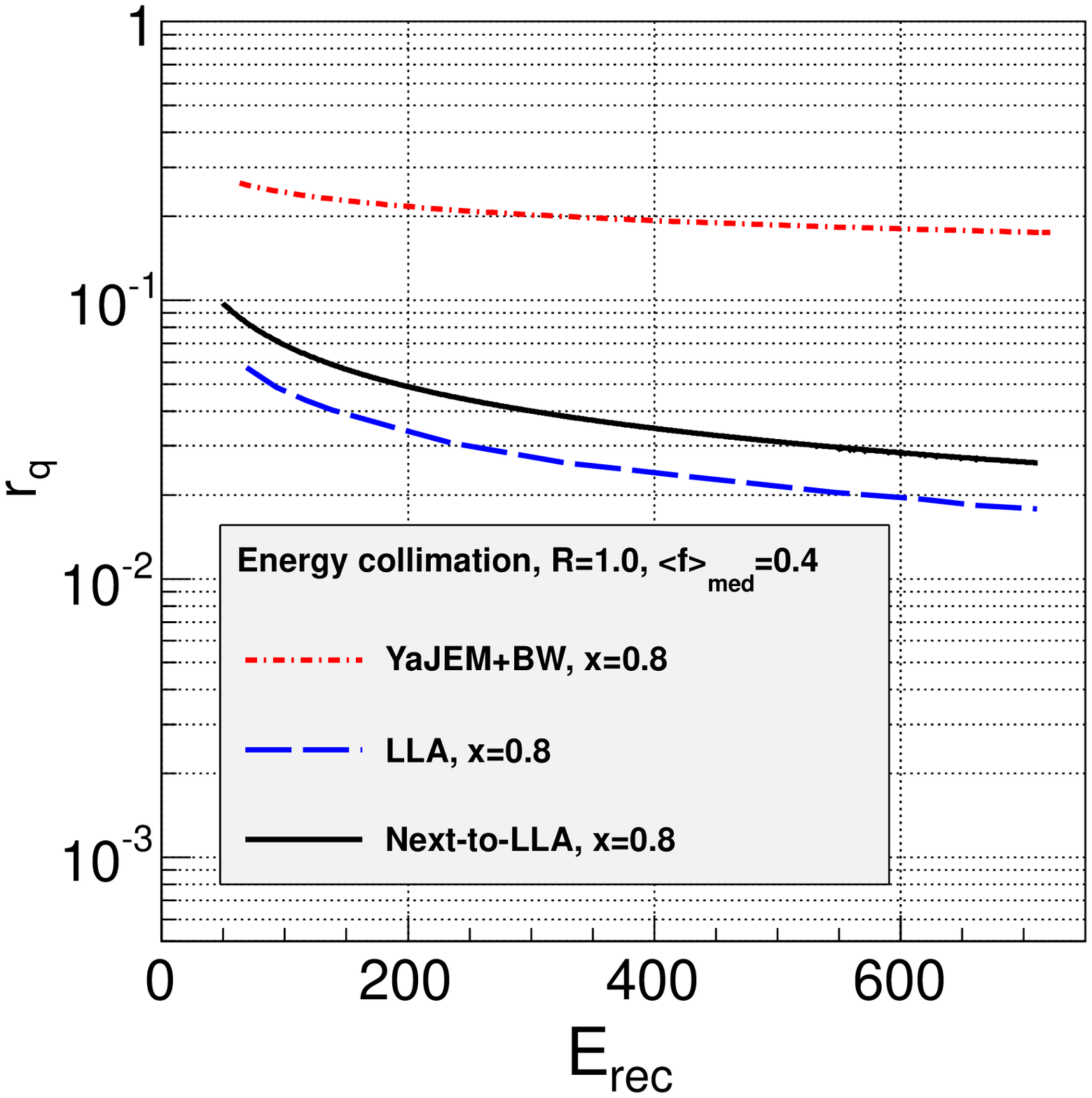, height=6.5truecm,width=7.8truecm}
\caption{\label{fig:collimquark} Collimation of energy inside a gluon jet for 
$x=0.5$ (left) and $x=0.8$ (right) with $R=1.0$.}  
\end{center}
\end{figure}

\begin{figure}[h]
\begin{center}
\epsfig{file=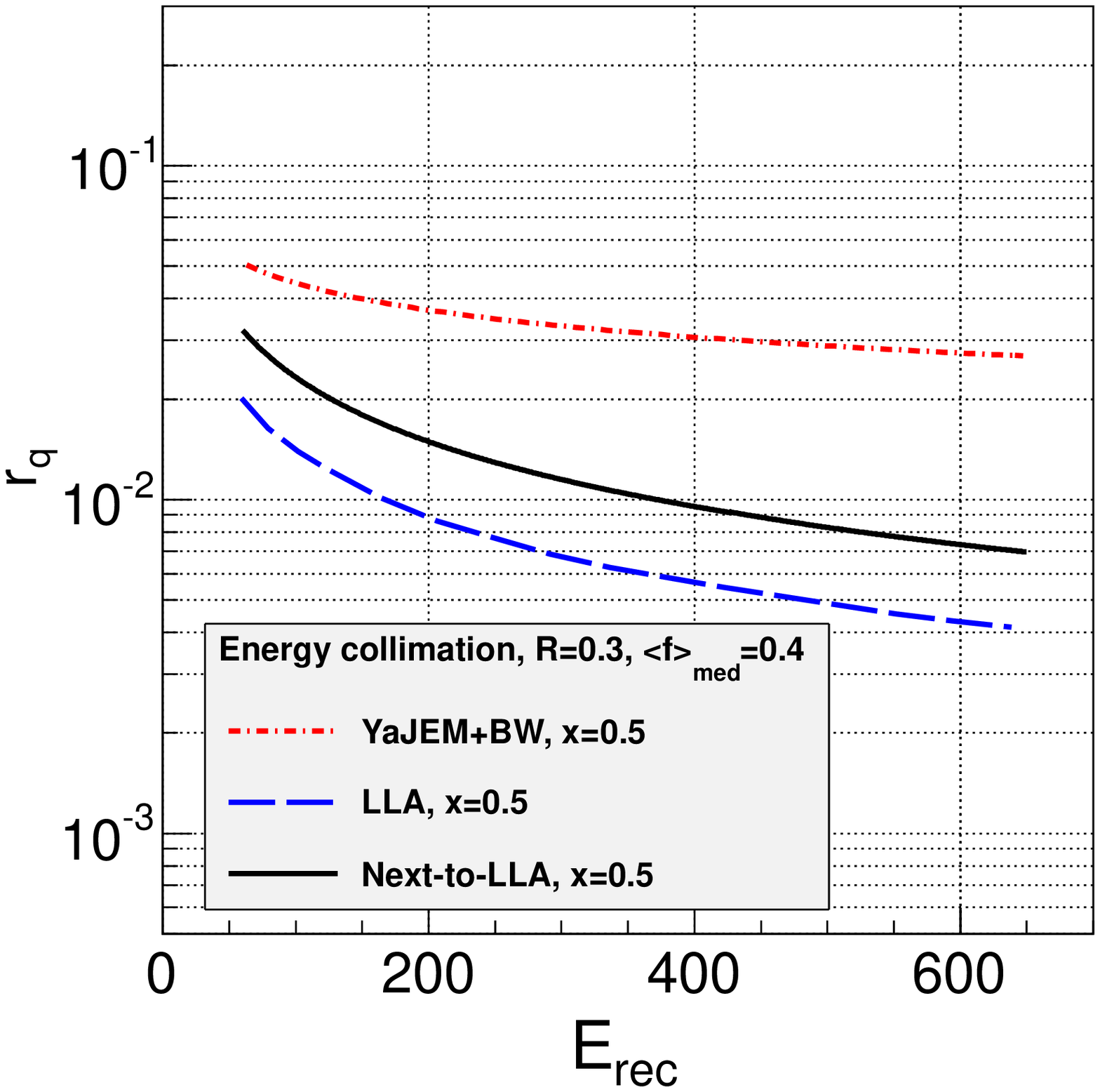, height=6.5truecm,width=7.8truecm}
\epsfig{file=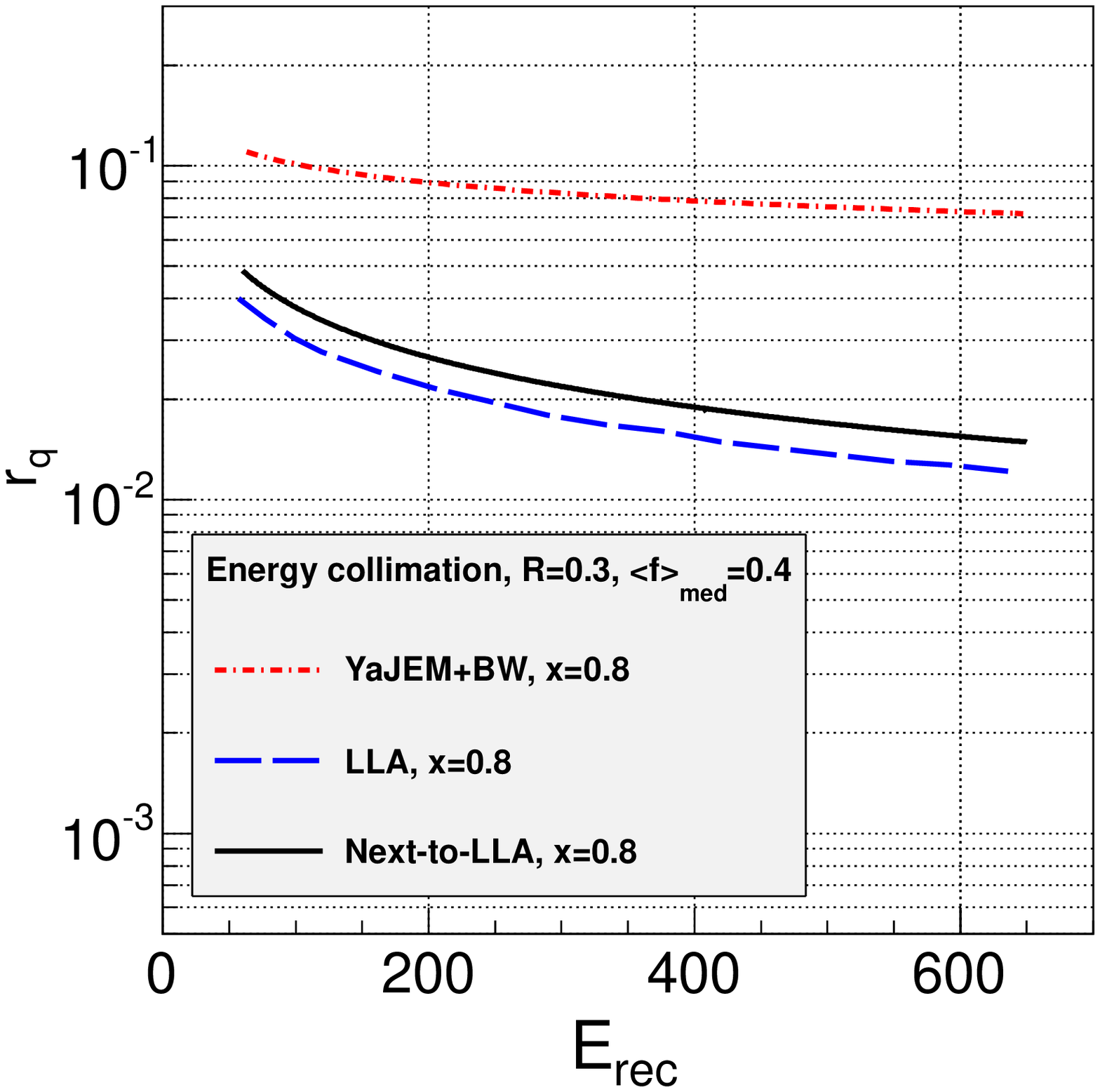, height=6.5truecm,width=7.8truecm}
\caption{\label{fig:collimquarkR03} Collimation of energy inside a quark jet for 
$x=0.4$ (left) and $x=0.8$ (right) with $R=0.3$.}  
\end{center}
\end{figure}

\begin{figure}[h]
\begin{center}
\epsfig{file=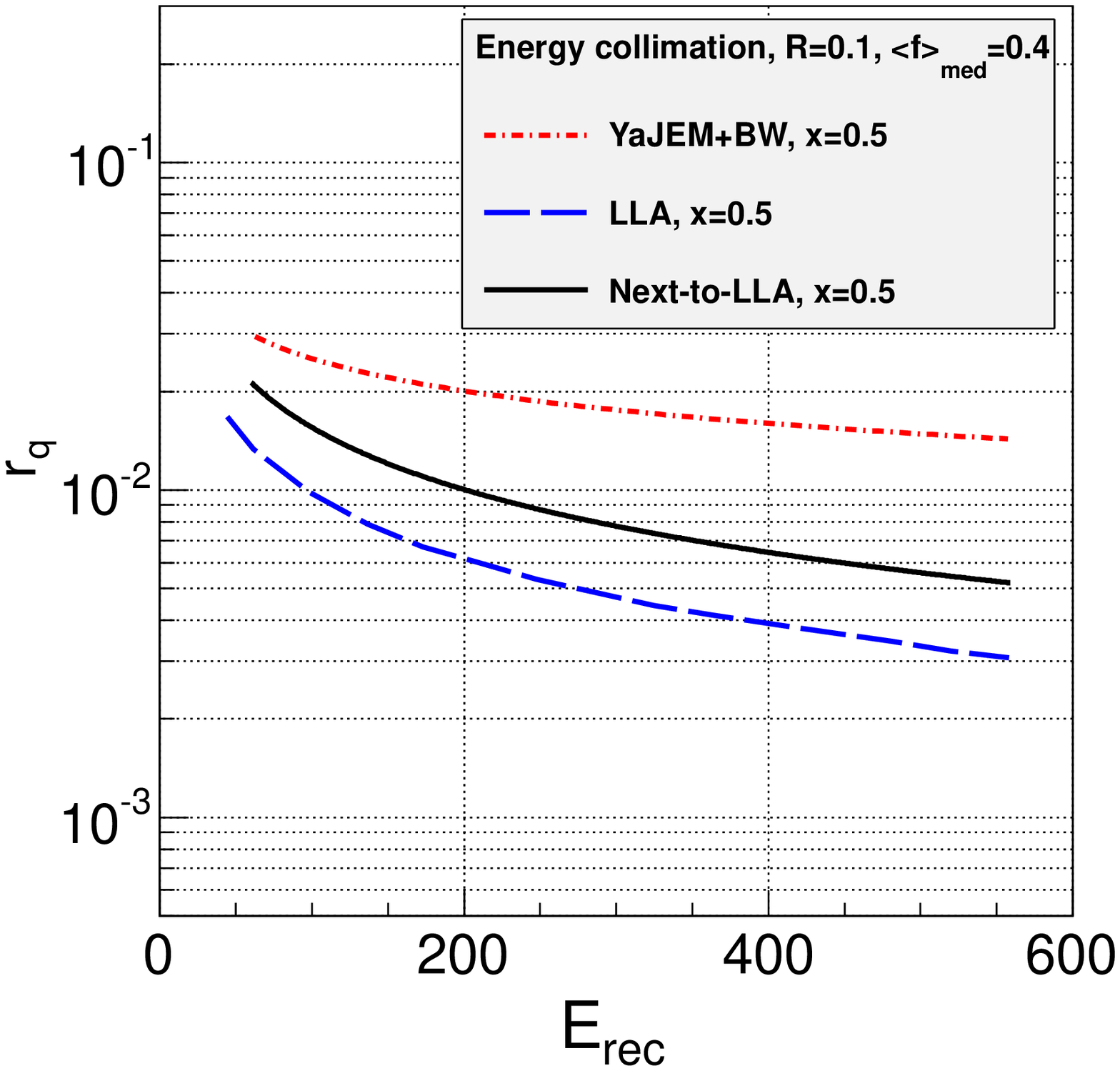, height=6.5truecm,width=7.8truecm}
\epsfig{file=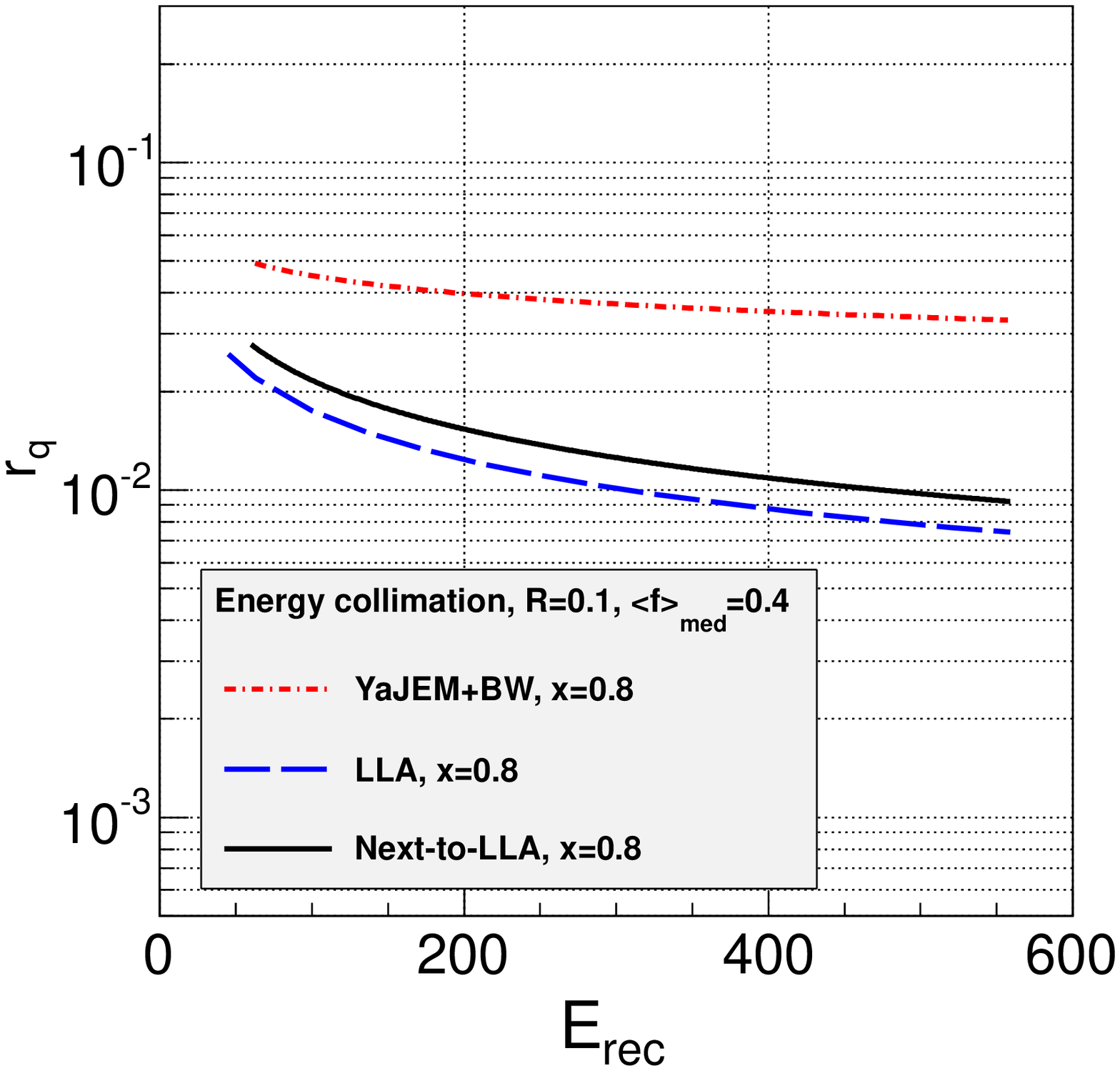, height=6.5truecm,width=7.8truecm}
\caption{\label{fig:collimquark01} Collimation of energy inside a gluon jet for 
$x=0.5$ (left) and $x=0.8$ (right) with $R=0.1$.}  
\end{center}
\end{figure}

\begin{figure}[h]
\begin{center}
\epsfig{file=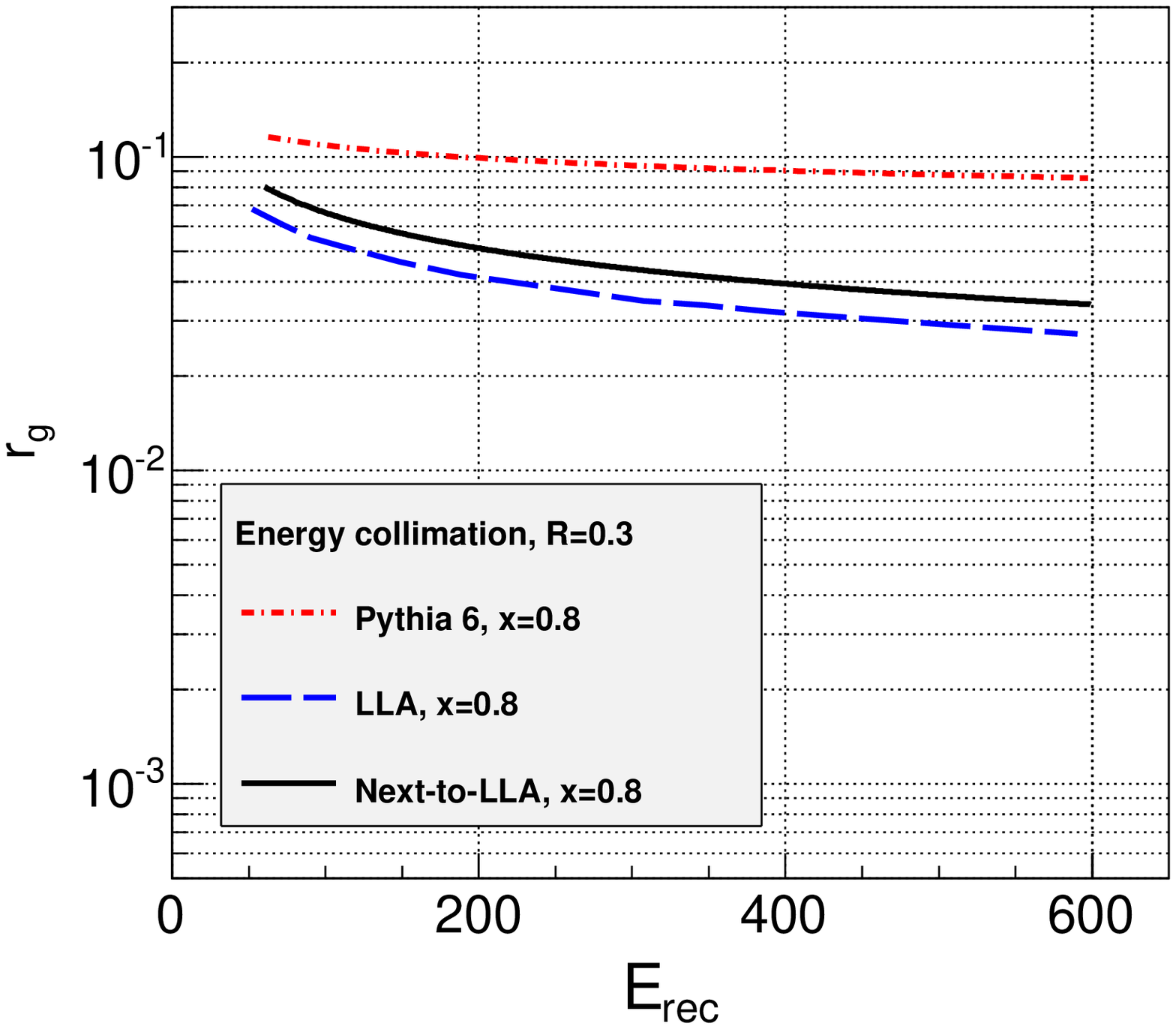, height=6.5truecm,width=7.8truecm}
\epsfig{file=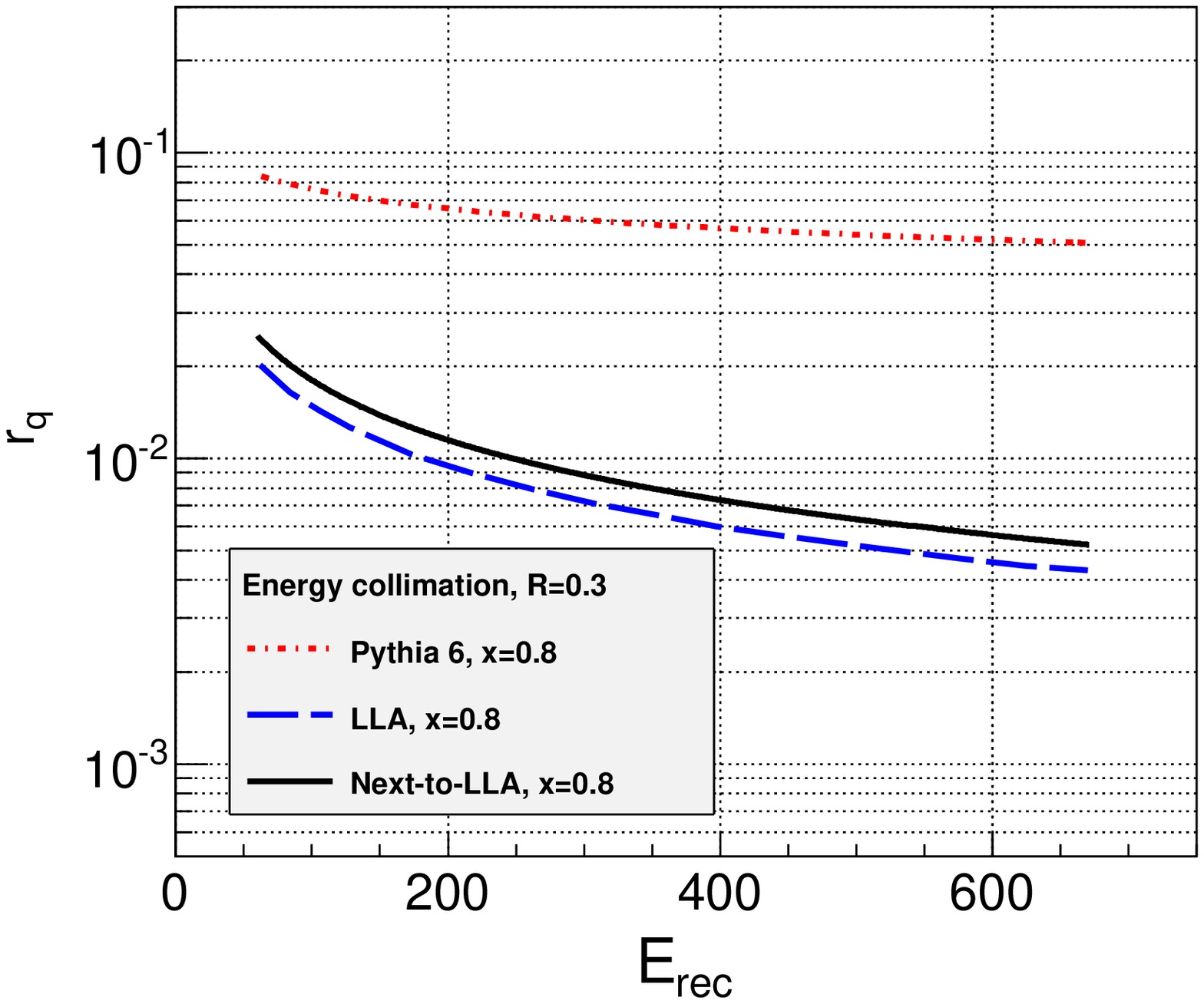, height=6.5truecm,width=7.8truecm}
\caption{\label{fig:collimgluonquarkvacuumR03} Collimation of energy inside a gluon jet (left) 
and quark jet (right) for $x=0.8$ with $R=0.3$ in the vacuum ($N_s=1$).}  
\end{center}
\end{figure}

In Table~\ref{table:reccoeffq}, we display quark dijets for the same values of 
center-of-mass energy and $R$. We can see that the 
recovered jet energy slightly increases compared with that displayed in 
Table~\ref{table:reccoeff} for gluon jets.
\begin{table}[htb]
\begin{center}
\begin{tabular}{cccccccc}
\hline\hline
$\sqrt{s}$ (GeV) & $R$
& $E_{\rm rec}$ (GeV) & $Q$ (GeV) & $R$ & $E_{\rm rec}$ (GeV) & $Q$ (GeV)\\ \hline
150 & 1.0 & 70.0 & 70.0 & $0.3$ & $58.5$ &  $17.6$ &\\
300 & 1.0 & 141.0 & 141.0 & $0.3$ & $100.1$ &  $30.0$ &\\
500 & 1.0 & 236.0 & 236.0 & $0.3$ & $205.9$ &  $61.5$ &\\
\hline\hline
\end{tabular}
\caption{Reconstructed jet energies inside the cone radii $R=1.0$ and $R=0.3$.}
\label{table:reccoeffq}
\end{center}
\end{table}
The energy collimation inside quark jets (\ref{eq:nscollim}) can be rewritten in the form,
\begin{equation}\label{eq:nscollim2}
r_q=R\left(\frac{Q}{\lqcd}\right)^{-\gamma_q(x,N_s)}.
\end{equation}
Accordingly, in Fig.\ref{fig:collimquark} and Fig.\ref{fig:collimquarkR03}, we display
the quark jet energy collimation for the energy fractions $x=0.5$ and $x=0.8$  
by the sub-jet with the medium modification value $\langle f\rangle_{\rm med}=0.4$. 
As for gluon jets, our predictions are in better agreement with YaJEM+BW for $R=0.3$ than for 
$R=1.0$. For $R=1.0$, the NLLA predictions underestimate YaJEM+BW for $x=0.5$
and $x=0.8$. However, for $R=0.3$, the disagreement is reduced as for gluon jets.
Furthermore, the correction due to the shift in $\ln x$ is smaller in a quark jet
compared to a gluon jet and LLA predictions in gluon jets 
are in better agreement with YaJEM+BW than in quark jets. The last statements suggest that NLLA and 
LLA predictions should be in better agreement with YaJEM+BW for much harder jets, i.e. $R=0.1$
as displayed in Fig.~\ref{fig:collimquark01}. 
The study of smaller jet resolutions such as $R=0.1$ which further biases QCD showers, 
is shown to always improve quark/gluon tagging at the LHC \cite{Gallicchio:2012ez}. 
\begin{figure}[h]
\begin{center}
\epsfig{file=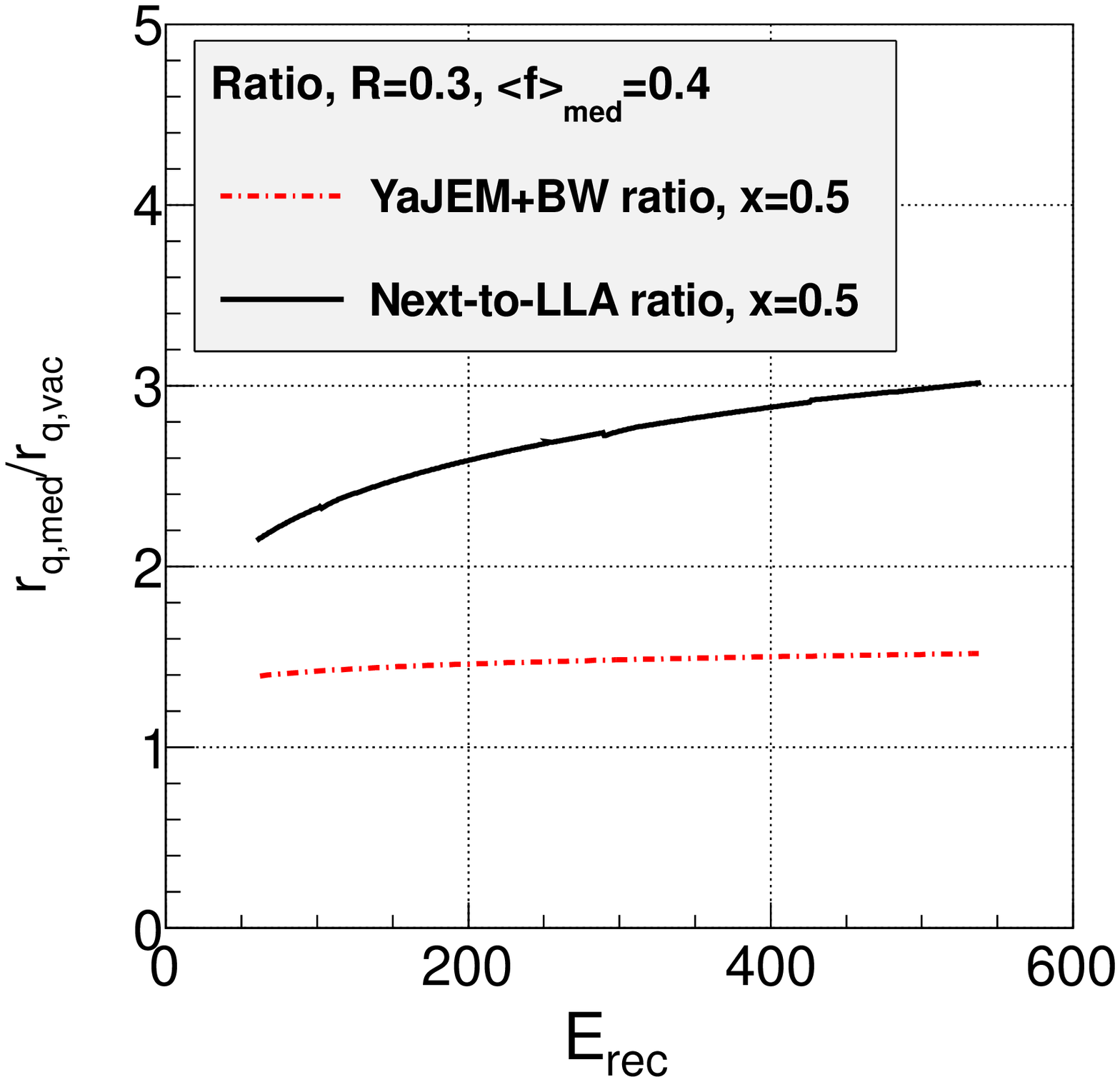, height=5.0truecm,width=5.2truecm}
\epsfig{file=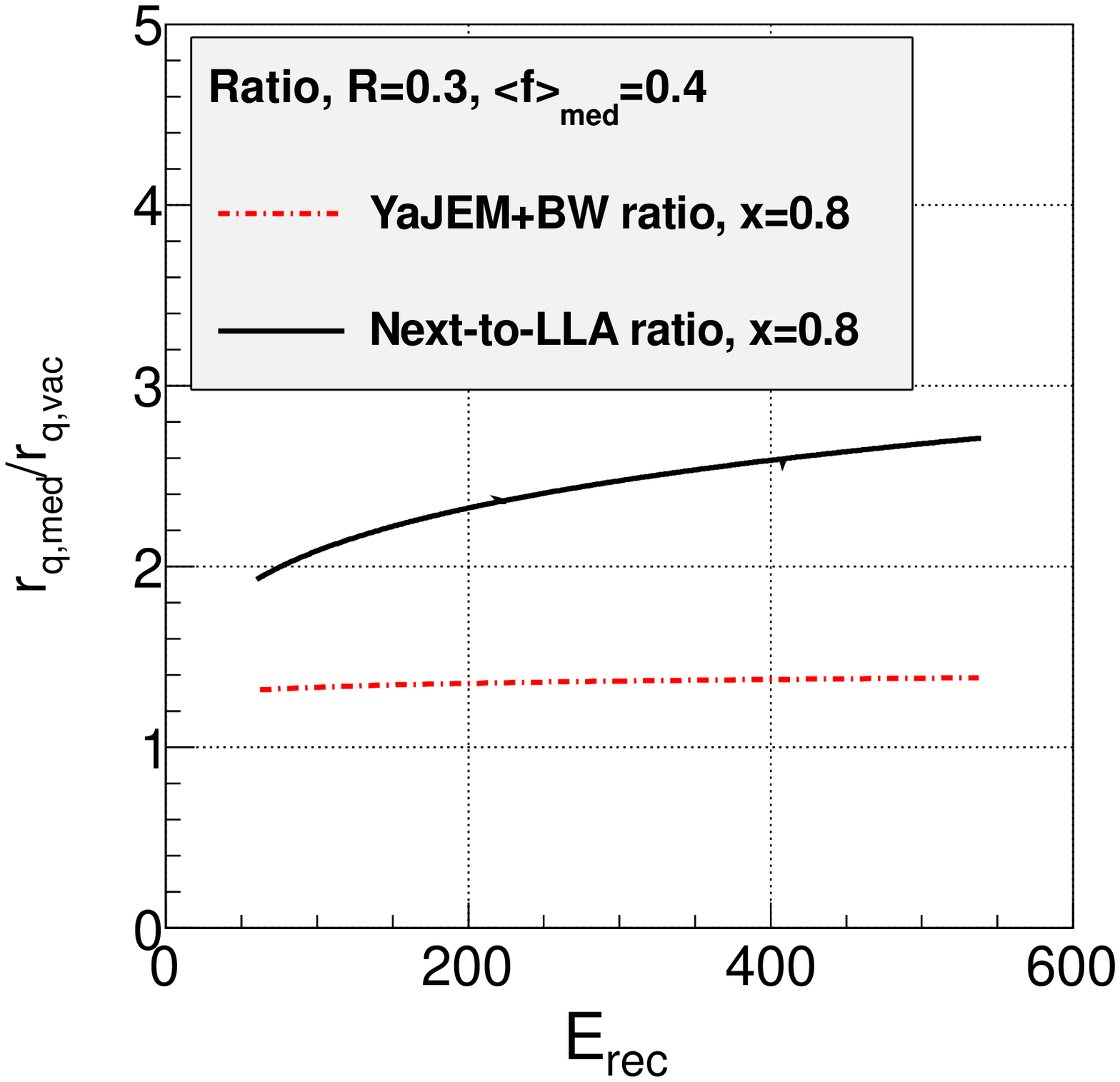, height=5.0truecm,width=5.2truecm}
\epsfig{file=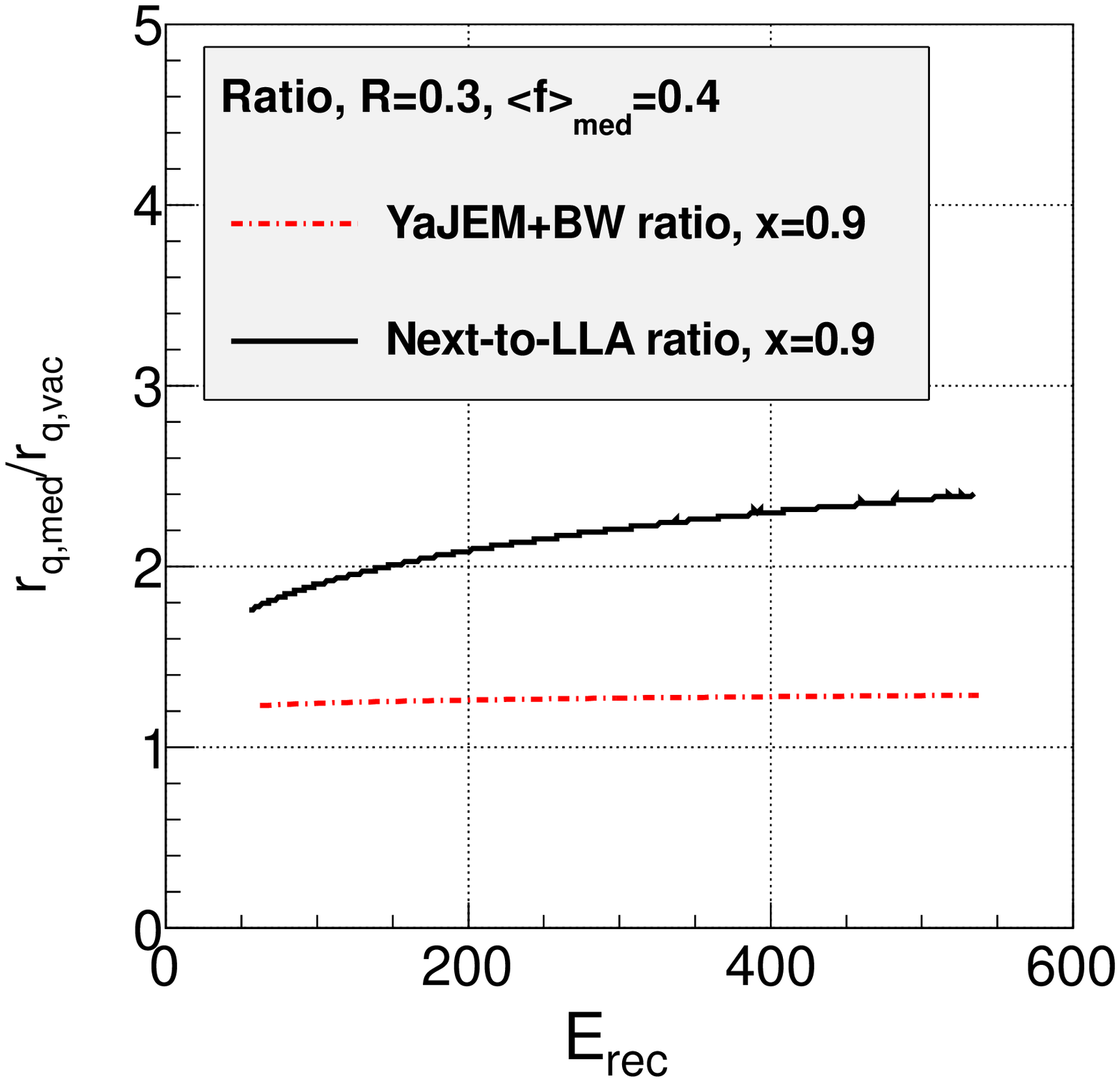, height=5.0truecm,width=5.2truecm}
\caption{\label{fig:ratiomedvsvacR03_Q} Medium-modified and vacuum energy collimation ratios 
$r_{q,\rm med}/r_{q,\rm vac}$ for $x=0.5$ (left), $x=0.8$ (center) and $x=0.9$ (right) with $R=0.3$.}  
\end{center}
\end{figure}
In Fig.~\ref{fig:ratiomedvsvacR03_Q}, we display the ratios $r_{q,\rm med}/r_{q,\rm vac}$ 
of the energy collimation in the medium and the vacuum. The results clearly show 
the quark jet broadening as a consequence of jet 
quenching. The comparison between the NLLA and LLA predictions is worse than for 
gluon jets. In Fig.\ref{fig:collimgluonquarkvacuumR03}, we compare the LLA, NLLA 
and Pythia 6 energy collimation in the vacuum.
The disagreement between the LLA, NLLA and Pythia 6 predictions is more pronounced in the vacuum, which further explains 
the huge difference displayed by the ratios in Fig.\ref{fig:ratiomedvsvacR03} and 
Fig.\ref{fig:ratiomedvsvacR03_Q}.

In Table~\ref{table:gammaAMC}, we present the values of the slopes $\gamma_A(x,N_s)$ provided
by YaJEM+BW ($\langle f\rangle_{\rm med}=1.4$) and Pythia 6. As displayed in the above figures for the energy collimation, 
the values are smaller than in the NLLA and LLA calculations presented in Table~\ref{table:gammaA}. 
However, the trends shown by the variation of $\gamma_A(x,N_s)$ as a 
function of $x$ and $N_s$ are similar ($\gamma_q>\gamma_g$). In particular, the slopes decrease as the energy fraction 
decreases for a given value of $N_s$. For a fixed value of $x$, the energy collimation flattens
as $N_s$ increases.
\begin{table}[htb]
\begin{center}
\begin{tabular}{ccccccc}
\hline\hline
YaJEM+BW & $x=0.5$ & $x=0.8$ & Pythia 6 & $x=0.5$ & $x=0.8$ \\ \hline
$\gamma_g(x,1.4)$ & 0.17 & 0.11  & $\gamma_g(x,1)$ & 0.21 & 0.14  &\\
$\gamma_q(x,1.4)$ & 0.24 & 0.17 & $\gamma_q(x,1)$ & 0.29 & 0.20 &\\
\hline\hline
\end{tabular}
\caption{YaJEM+BW values of the slope $\gamma_A(x,N_s)$ of the energy collimation 
for $N_s=1.4$ (medium) and Pythia 6 values for $N_s=1$ (vacuum).}
\label{table:gammaAMC}
\end{center}
\end{table}
We can see that our calculations and YaJEM+BW
predict a much stronger energy collimation in quark jets than in gluon jets. 
Physically, this is because gluon jets have a color charge roughly twice as 
large ($N_c/C_F=9/4$) than quark jets, or equivalently, gluon jet multiplicities
are higher than quark jet multiplicities by the same factor asymptotically \cite{Mueller:1983cq}. 
Moreover, the first splitting dominates the jet width, which for quark jets 
only has the available splitting $q\to qg$ where the emitted gluon is preferentially soft and does not 
alter the transverse jet shape, whereas gluon jets can split into $g\to q\bar q$ pairs 
where both quarks tend to be equally hard, which can widen the shape substantially. 
In both YaJEM+BW and the calculation, the energy collimation is hence steeper for quark jets 
than for gluon jets.

Equations (\ref{eq:newcollimformbis}) and (\ref{eq:digammaeq}) provide a simple description  
of the jet energy collimation under consideration and cannot be in perfect agreement with 
the YaJEM+BW description. The last point suggests that other perturbative contributions arising from
the splittings $q\to qg$ and $g\to q\bar{q}$ should be included in Eq.~(\ref{eq:sumDAB}) 
in the form $D_q(x,E\Theta_0,xE\Theta)=D_q^q(x,E\Theta_0,xE\Theta)+D_q^g(x,E\Theta_0,xE\Theta)$
for quark jets and $D_g(x,E\Theta_0,xE\Theta)=D_g^g(x,E\Theta_0,xE\Theta)+D_g^q(x,E\Theta_0,xE\Theta)$
for gluon jets, with the full resummed contribution of the 
soft-collinear logarithms in DGLAP FFs \cite{Albino:2004xa,Albino:2009hu}. 
Indeed, as the jet energy increases, the contributions from the double logarithmic contributions 
$\alpha_s\frac{dz}{z}\frac{d\Theta}{\Theta}$  ($z=E_g/E$) increase asymptotically, which may explain 
why the difference between YaJEM+BW and the NLLA predictions gets wider as the jet energy 
$E$ increases. 
Moreover, the  more accurate treatment of phase space 
in both Pythia and YaJEM have not been taken 
into account in the NLLA and LLA calculations.

\subsection{Hadronization effects in the energy  collimation}

\begin{figure}[h]
\begin{center}
\epsfig{file=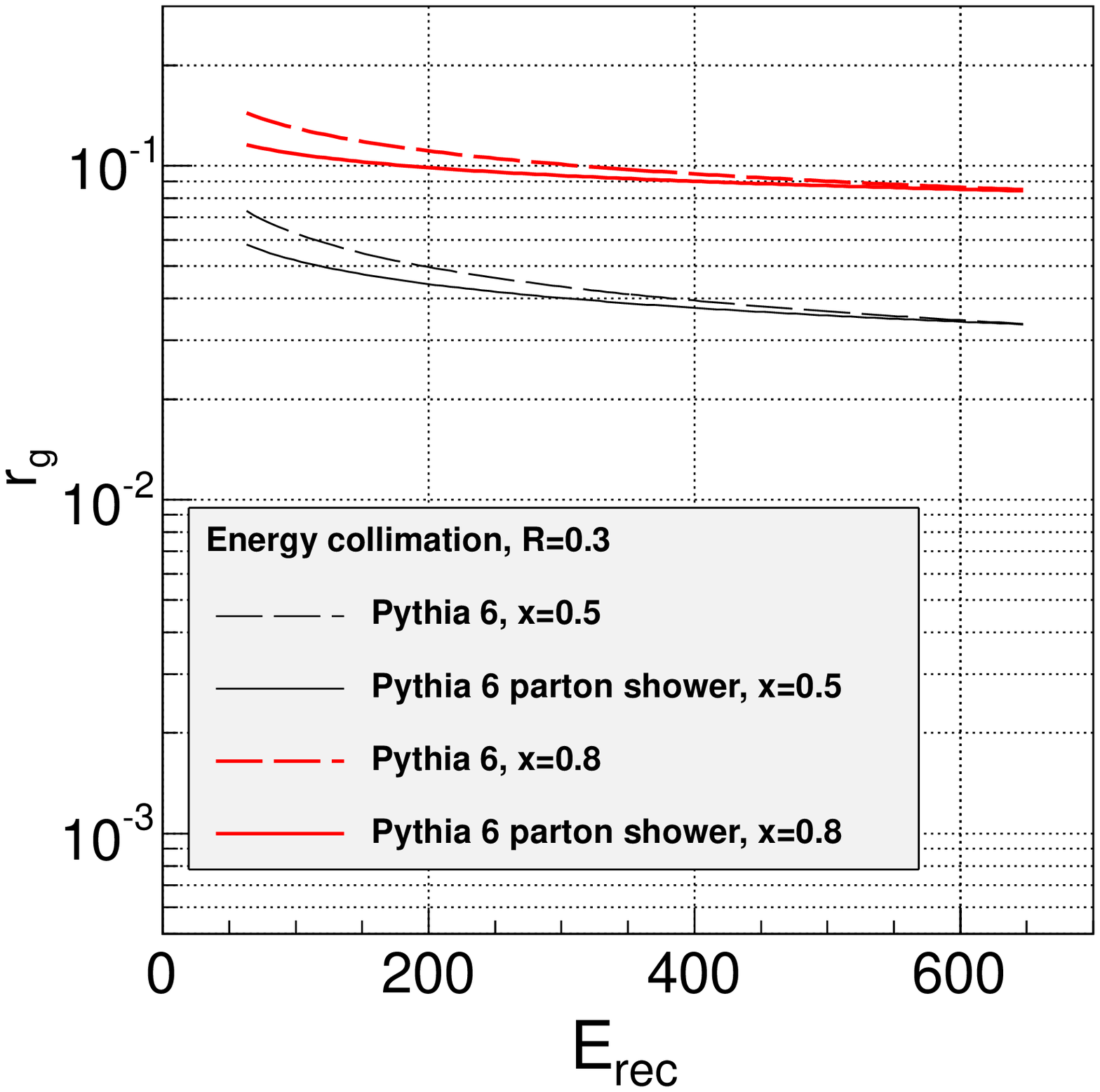, height=6.5truecm,width=7.8truecm}
\epsfig{file=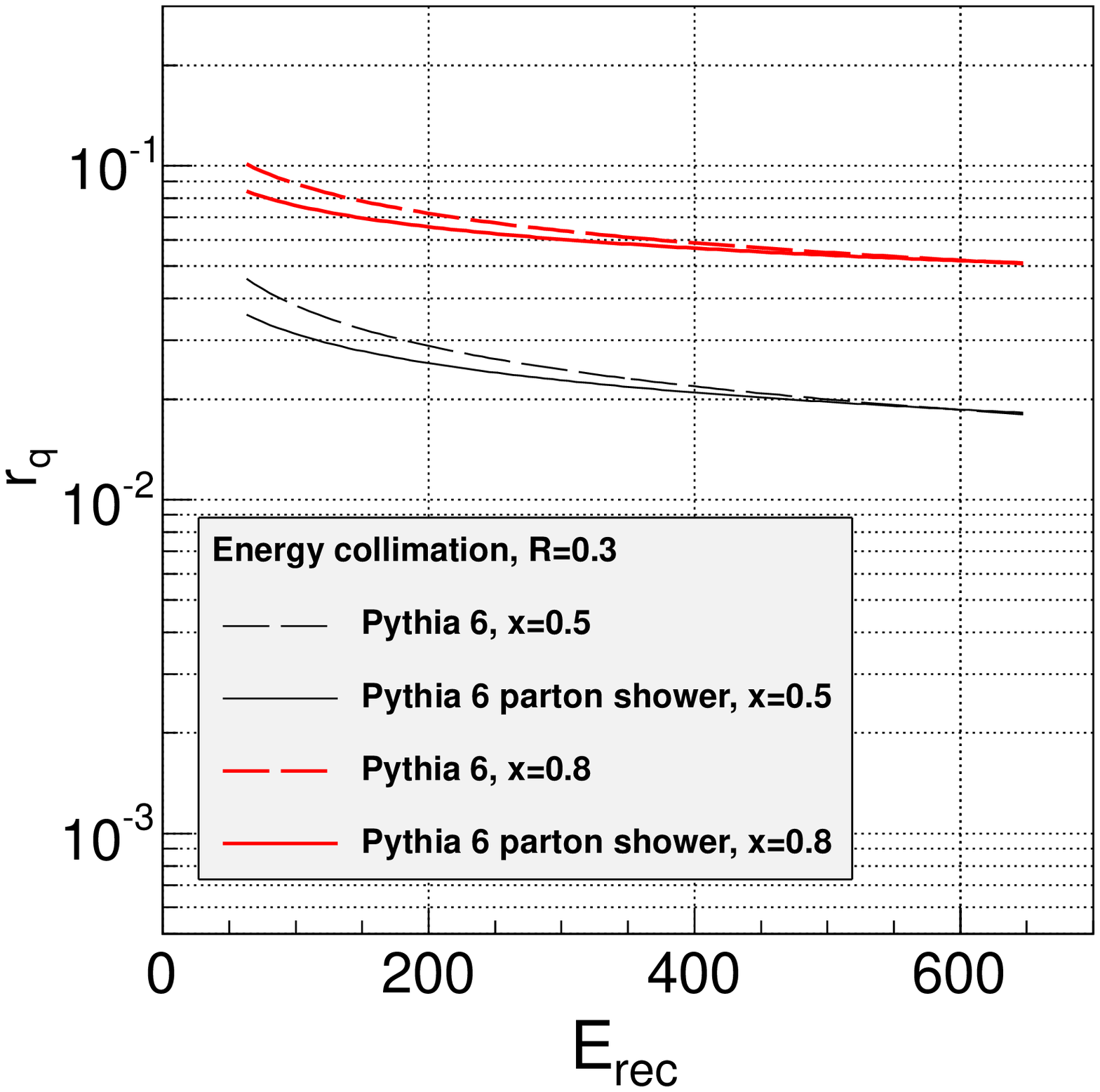, height=6.5truecm,width=7.8truecm}
\caption{\label{fig:partonvshadron} Parton versus hadron energy collimation  
for $x=0.5$ and $x=0.8$ inside a gluon jet (left) and a quark jet (right)  with $R=0.3$.}  
\end{center}
\end{figure}

In Fig.~\ref{fig:partonvshadron} we display the energy collimation inside gluon and quark 
jets in the vacuum using Pythia 6. The role of hadronization is displayed by 
comparing the energy collimation for final-state hadrons and final-state partons clustered
inside the radius $R=0.3$ in the energy range $50\leq E_{\rm rec}(\rm GeV)\leq700$ by using
the anti-$k_T$ algorithm \cite{Cacciari:2011ma,Cacciari:2005hq}. 
The hadronic energy collimation has been labeled as ``Pythia 6" and the partonic 
energy collimation, ``Pythia 6 parton shower." Since the hadronization is modeled to occur outside the medium, 
we limited the comparison to Pythia 6 since the results would be identical 
for YaJEM+BW with a slightly larger normalization (see Figs.~\ref{fig:collimgluonR03} 
and \ref{fig:collimquarkR03} for comparison) as a consequence of the jet 
broadening.
As we can see, the hadronization biases the partonic energy collimation for jet energies
$<400$ GeV but this effect is $\sim5\%$ at RHIC energy scales and smaller than $1\%$
at LHC energy scales. For jet energies $>400$ GeV the role of hadronization becomes 
negligible and therefore irrelevant for the study of this observable. In general
hadronic showers are less collimated than a fictitious parton shower at small
energy scales. 

\section{Jet shape: comparison with PbPb CMS data}
\label{section:CMSdatacomparison}
The integrated jet shape $\Psi(r;R)$ measures the fraction of the jet energy of size $R$
contained in a sub-cone of size $r$ such that $\Psi(R;R)=1$. The differential jet shape 
reads \cite{Seymour:1997kj},
$$
r\psi(r;R)=r\frac{d\Psi}{dr},
$$
where
\begin{equation}\label{eq:diffjetshape}
\frac{d\Psi}{dr}=\frac1{E_{rec}}\int de\,e\frac{d^2N}{dedr'},
\quad E_{\rm rec}=\int_0^Rdr'\int de\,e\frac{d^2N}{dedr'}.
\end{equation}
Hence, the integration of Eq.~(\ref{eq:diffjetshape}) leads to the expression written in 
Eq.~(\ref{eq:yajemcollim}) for the energy fraction $x$ used in the framework of the energy collimation, 
which will be identified with the integrated jet shape hereafter: $x\equiv\Psi(r;R)$. Our NLLA and 
LLA predictions for the integrated jet shape will be therefore based on the maximal angular 
aperture $\Theta\equiv r$ where the bulk of the jet energy is contained, as we discussed in 
Sec.~\ref{sec:theory}. For the first time, in this paper the jet shape is computed from the 
jet energy collimation within the same NLLA and LLA schemes. 
 
First of all, we describe how the Monte Carlo simulation from Pythia 6 and YaJEM 
is performed in view of further comparison with LLA, NLLA predictions and CMS
data hereafter. 

For the computation of the integrated jet shapes extracted 
from (\ref{eq:diffjetshape}), we will limit our study to charged particles only,  
as in the CMS experiment \cite{Chatrchyan:2013kwa}.
The initial distribution of gluon- and quark-initiated showers for the analysis
is determined by the convolution of PDFs and nPDFs with the LO matrix elements of the 
final cross section at the given hard factorization scale of the process. The LO matrix elements
of the partonic cross section can be computed analytically \cite{Eskola:2002kv}.
PDFs and nPDFs are provided by the CTEQ \cite{Lai:2010nw} 
and EKS \cite{Eskola:1998df} families for $pp$ and PbPb collisions in the vacuum and the medium respectively.
The analysis carried out for the jet shapes is hence different than the analysis for the
energy collimation in Secs.~\ref{subsec:gluoncollim1} and \ref{subsec:quarkcollim1} where
the center-of-mass energy of the hard parton system was fixed to a certain value $\sqrt{s}$.

A large number of quark and gluon dijets are randomly generated based on the 
perturbative QCD spectrum inside the energy range $200\leq \sqrt{s}({\rm GeV})\leq 600$ and clustered by using the  
anti-$k_t$ algorithm with $R=0.3$. After clustering all charged hadrons 
with $e>1$ GeV inside the
given cone $R=0.3$, jet energies $E_{\rm rec}$ are required to fulfill CMS trigger conditions 
imposed by the restriction $E_{\rm rec,jet}\geq 100$ GeV. 
The requirement imposed by the trigger selection in the analysis
will be referred to as a biased shower, while that including all clustered jet energies will be 
referred to as an unbiased shower in the following. Accordingly, the fraction of gluon jets in one sample is
biased by the trigger condition from $f_g^{\rm vac}\approx0.4$ in the unbiased case to 
$f_g^{\rm vac}\approx0.2$ in the biased case in vacuum showers; 
and from $f_g^{\rm med}\approx0.3$ to $f_g^{\rm med}\approx0.1$ in medium showers. 
Thus, quark jets are dominant in the analysis,
particularly in the medium.
\begin{figure}[h]
\begin{center}
\epsfig{file=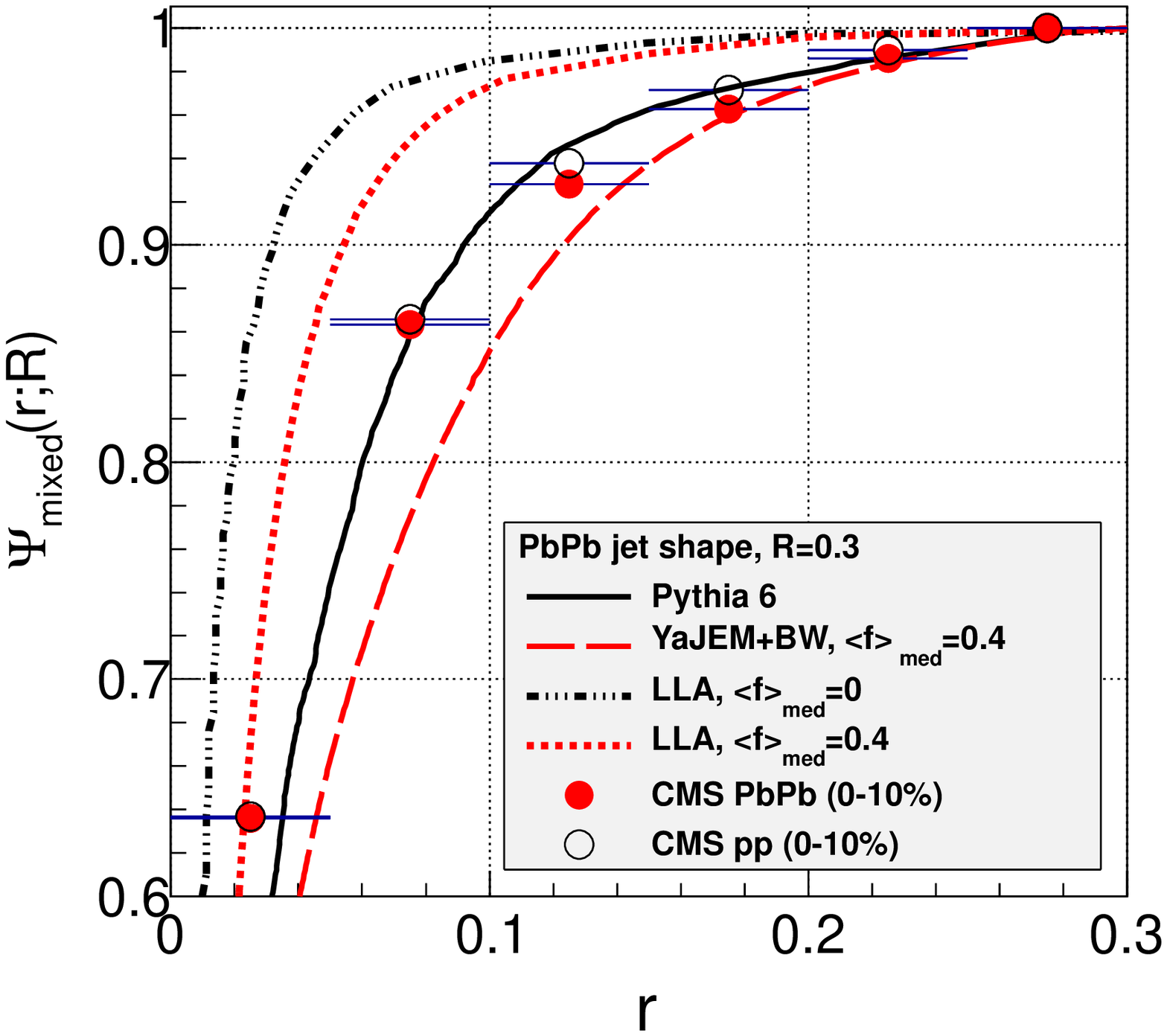, height=6.5truecm,width=7.8truecm}
\epsfig{file=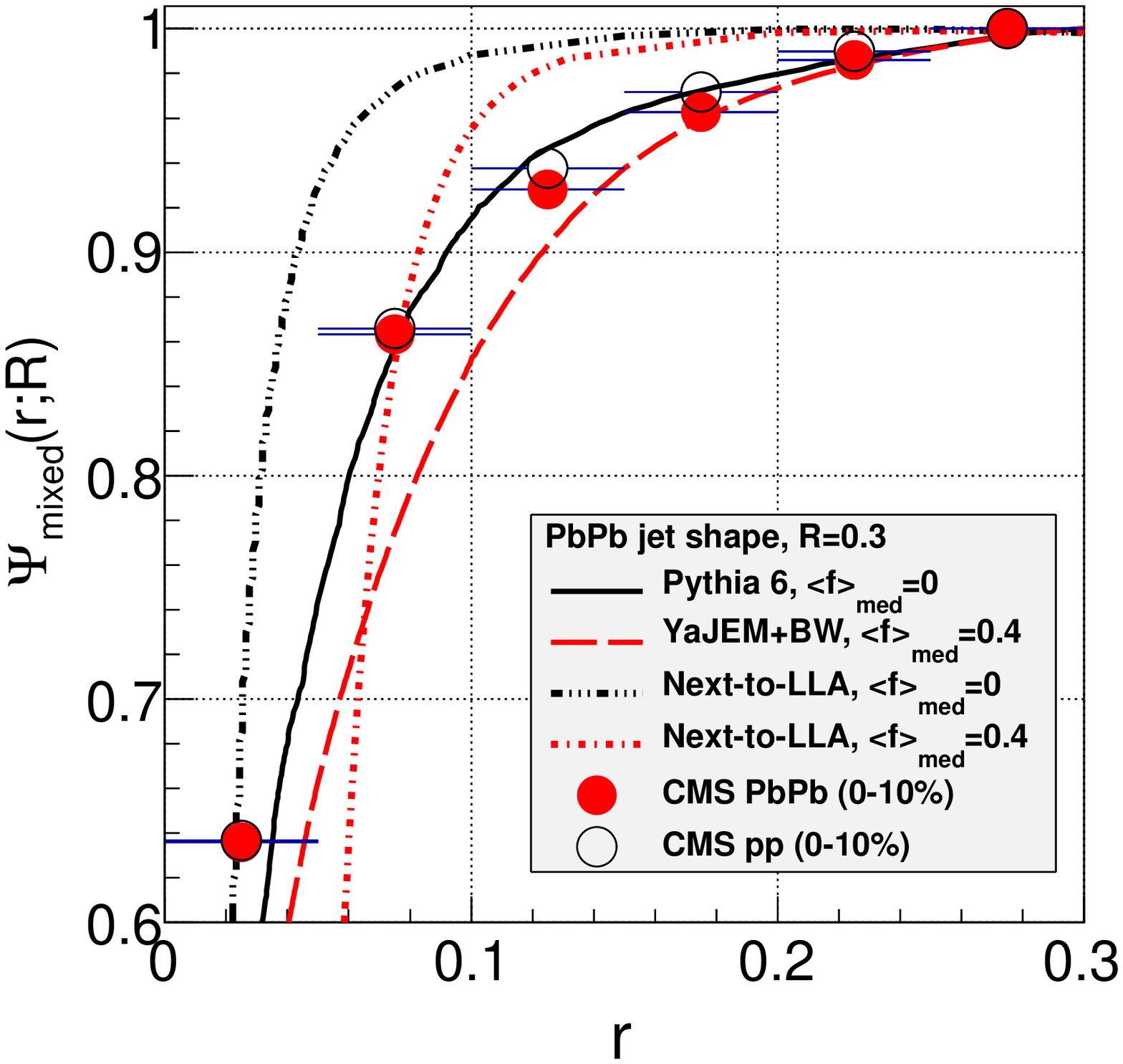, height=6.5truecm,width=7.8truecm}
\caption{\label{fig:jetshapePbPbppdata} Jet shape for CMS $pp$ and PbPb data with $R=0.3$, 
compared with Pythia 6, YaJEM+BW, the LLA formula (left panel) and NLLA formula (right panel).}  
\end{center}
\end{figure}
The fraction of gluon jets in a sample is used for the computation of the mixed 
integrated jet shape given by the linear combination for gluon and quark jets
in the form,
\begin{equation}\label{eq:mixedjetshape}
\Psi_{\rm mixed}(r;R)=f_g\Psi_g(r;R)+(1-f_g)\Psi_q(r;R)
\end{equation}
for a direct comparison of the LLA (\ref{eq:digammaeq}), the NLLA (\ref{eq:newcollimformbis}), 
Pythia 6 and YaJEM+BW with CMS data.

In YaJEM+BW, $f_{\rm med}$ is computed event by event as described in  
Sec.~\ref{subsec:yajemdescription}.
Instead of treating quark and gluon jets 
independently in this framework, it is more straightforward to mix the differential
distributions for the energy flux (\ref{eq:diffjetshape}) and compare them with the mixed jet 
shape from Pythia and YaJEM+BW with account of hadronization. 
By doing so, after averaging over a large number of events, all jets cluster to the biased mean 
jet-energy value $E_{\rm rec}\sim140$ GeV and the mean medium parameter $\langle f\rangle_{\rm med}\sim0.4$.

In order to compare the Pythia 6 and YaJEM+BW calculations with the 
LLA and NLLA gluon and quark jet shapes, we solve the equations 
(\ref{eq:newcollimformbis}) and (\ref{eq:digammaeq}) numerically 
[$x_g\equiv\Psi_g(r;R)$ and $x_q\equiv\Psi_q(r;R)$] as a function of $r$ in the interval
$0\leq r\leq R$ in the framework of LLA (NLLA) DGLAP evolution
at large $x$, for the first time in this paper. For the computation, we choose the mean jet-energy value 
$E_{\rm rec}=140$ GeV extracted from Pythia 6 and YaJEM+BW. Taking the same values for the 
fraction of gluon jets in a sample $f_g^{\rm vac}\approx0.2$ in the vacuum 
($\langle f\rangle_{\rm med}=0$) and $f_g^{\rm vac}\approx0.1$
in the medium ($\langle f\rangle_{\rm med}=0.4$), we can evaluate the mixed LLA and NLLA jet shapes 
(\ref{eq:mixedjetshape}) equivalently as in the Monte Carlo event generators.  
 
In Fig.~\ref{fig:jetshapePbPbppdata}, we show the LLA, 
Pythia 6 and YaJEM+BW for $\langle f\rangle_{\rm med}=0.4$
jet shapes compared with $pp$ and PbPb CMS data in the left 
panel, the NLLA, Pythia 6 and YaJEM+BW jet shapes for 
$\langle f\rangle_{\rm med}=0.4$ compared with $pp$ and 
PbPb CMS data in the right panel. The jet shape is displayed
in the interval $0.6\leq x\equiv\Psi(r;R)\leq1$ in agreement with
the LLA (and NLLA) DGLAP large sub-jet energy fraction $x$ approximation where these calculations
are performed. We can see that the LLA and NLLA qualitatively 
describe the features of the jet shapes in both vacuum and medium but an 
important disagreement persists. Compared to the LLA calculation, the NLLA calculation approaches the data as 
the jet shape decreases and particularly for more collimated sub-jets $r$, as expected.
Translating this statement to the energy collimation, we show the NLLA correction to widen 
the energy dependence of $r$ and to increase the difference with the LLA
calculation as the sub-jet energy fraction (jet shape) decreases. 
Thus, although the NLLA and LLA predictions seem to capture the main ingredients of the 
hard fragmentation process in this framework, the account of all fragmentation 
probabilities, mainly those containing soft gluon contributions should be taken into account in a 
more accurate theoretical framework. Pythia 6 provides instead a good agreement 
with $pp$ CMS data for biased jets. YaJEM+BW describes the shower medium modifications by
reproducing the jet broadening at slightly larger values of $r$ to the right, similar to
the medium-modified NLLA and LLA jet shapes. Though the YaJEM+BW
calculation does not reproduce the data points exactly, the curve fits 
the systematic error bars of the CMS PbPb data. As observed, the sub-jet broadening shown by the data is 
very small but in better agreement with the sub-jet broadening shown by the Monte Carlo simulations than with 
that shown by the theoretical calculations with the BW prescription. 
\begin{figure}[h]
\begin{center}
\epsfig{file=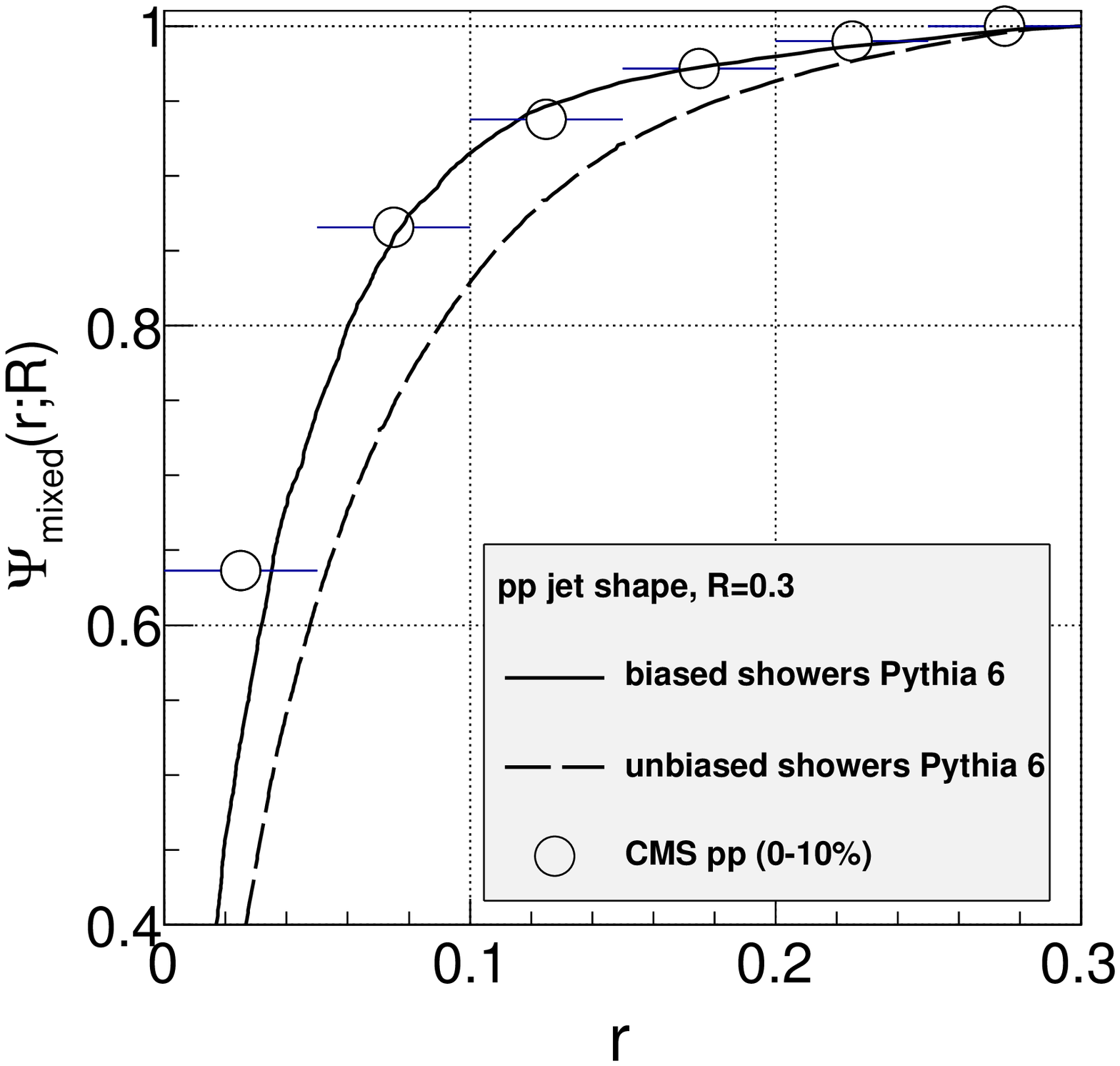, height=6.5truecm,width=7.8truecm}
\epsfig{file=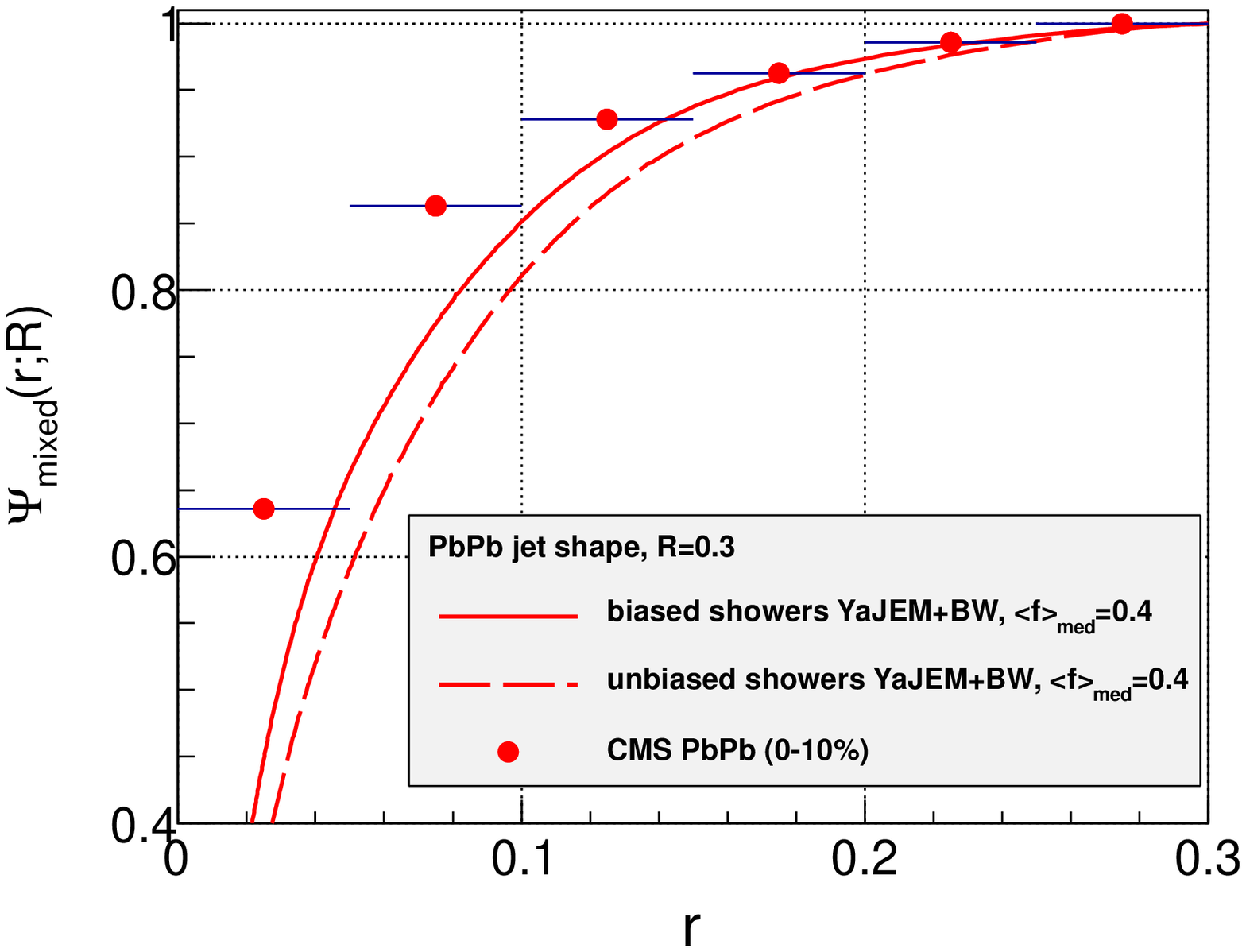, height=6.5truecm,width=7.8truecm}
\caption{\label{fig:biasedvsunbiased} Biased versus unbiased jet shape for CMS $pp$ data
(left panel) and PbPb data (right panel) with $R=0.3$.}  
\end{center}
\end{figure}

In Fig.~\ref{fig:biasedvsunbiased} we compare the biased ($E_{\rm rec,jet}\geq100$ GeV)
and unbiased (all jets) cases obtained with 
Pythia 6 and YaJEM+BW with $pp$ (left panel) and PbPb (right panel) CMS data. 
The unbiased mean jet energy turns out to be $E_{\rm jet}\sim90$ GeV after all
clustered jets are considered in the analysis without any further trigger bias. 
Furthermore, the evaluation of the unbiased 
case through Eq.~(\ref{eq:mixedjetshape}) requires the unbiased gluon jet fractions
$f_g^{\rm vac}\approx0.4$ and $f_g^{\rm med}\approx0.3$ in the vacuum and in the medium respectively, as performed here.
As can be seen, the shower structure is affected by imposing a jet-energy condition which leads
to a better agreement between the biased jet shape and the data than
the unbiased jet shape.

\subsection{Hadronization effects in gluon and quark jet shapes}
\begin{figure}[h]
\begin{center}
\epsfig{file=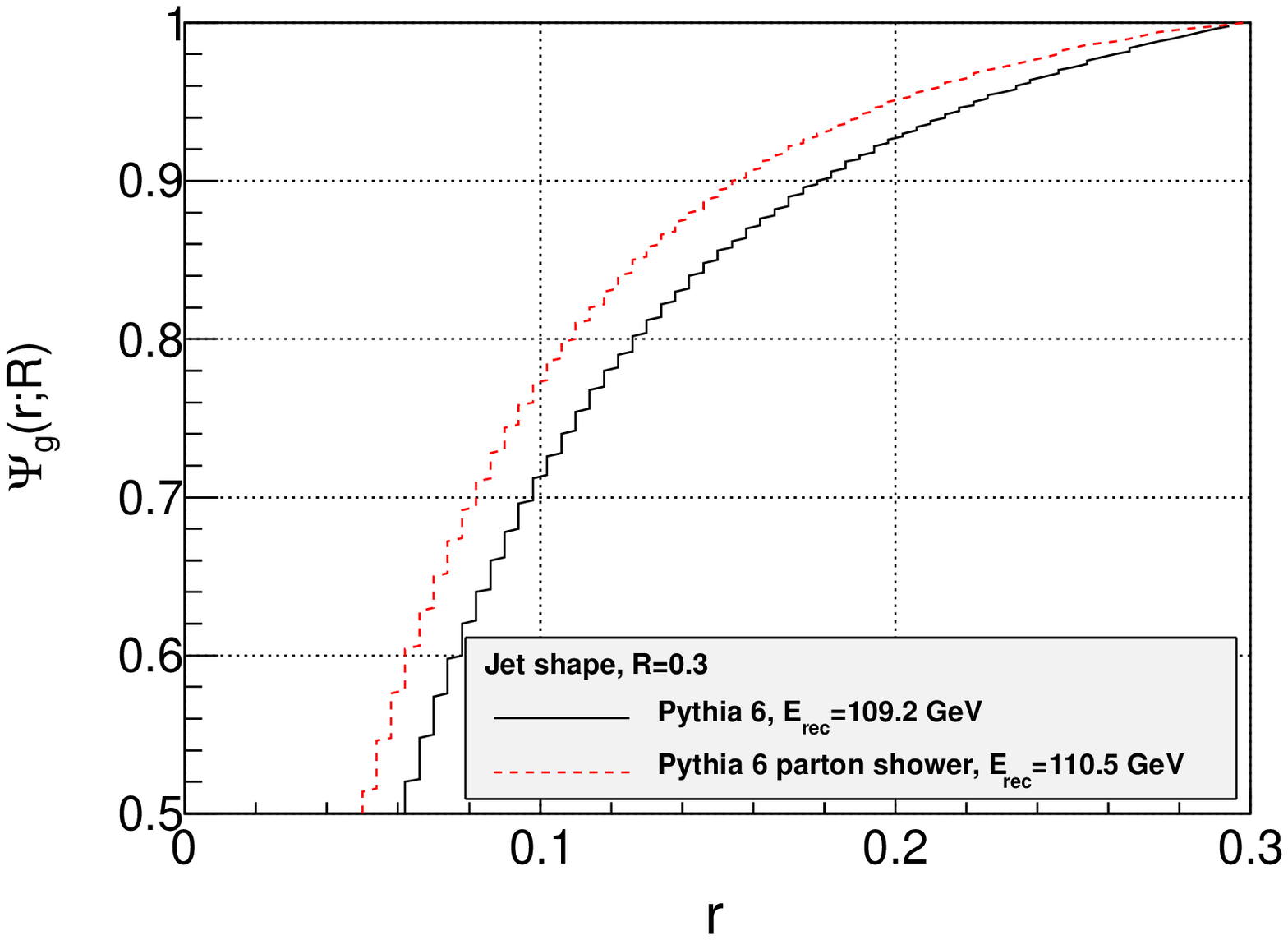, height=6.5truecm,width=7.8truecm}
\epsfig{file=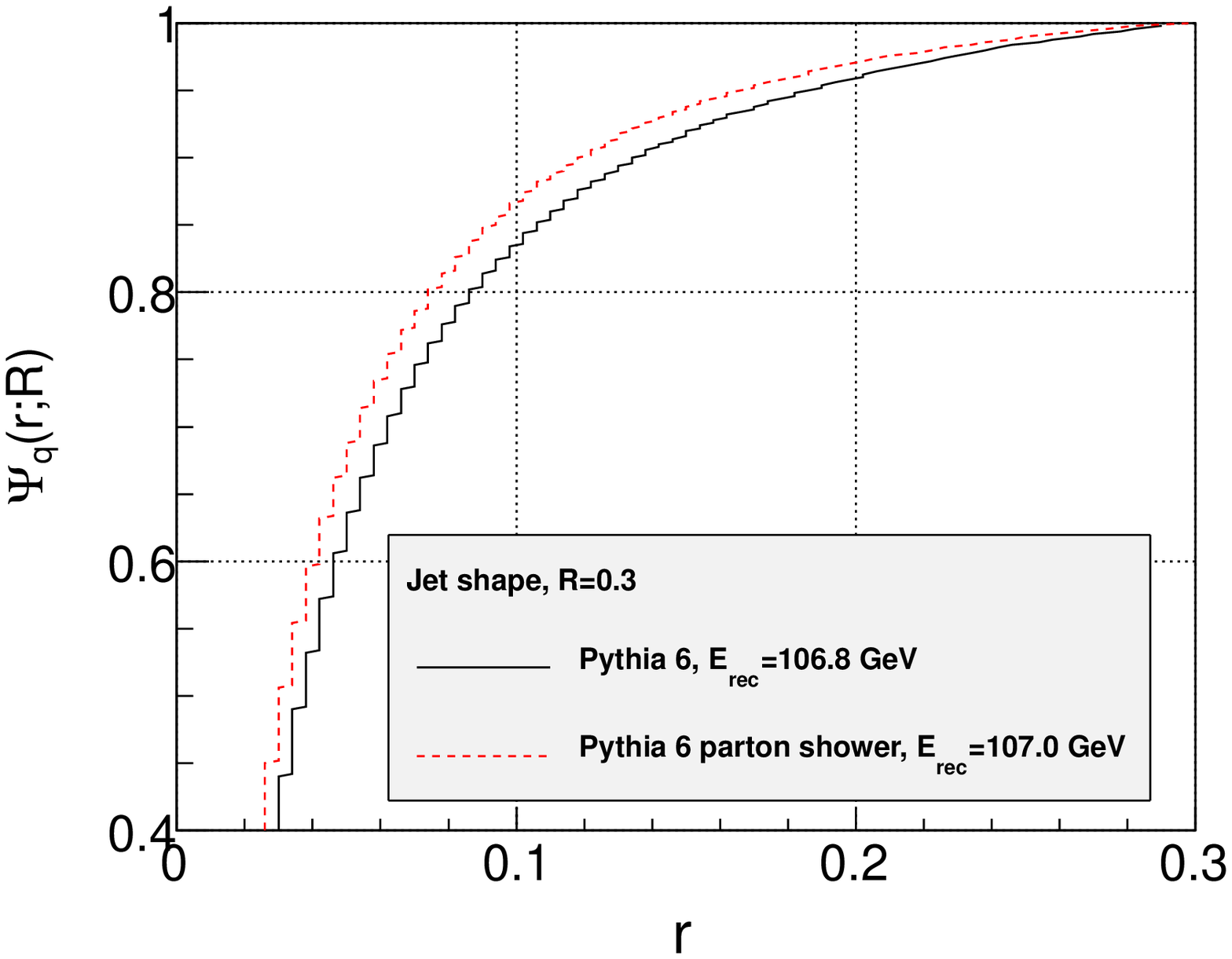, height=6.5truecm,width=7.8truecm}
\caption{\label{fig:partonvshadronjetshape} Parton versus hadron jet shape 
inside a gluon jet (left) and a quark jet (right)  with $R=0.3$.}  
\end{center}
\end{figure}

Finally, in Fig.~\ref{fig:partonvshadronjetshape}, we display the jet shape obtained 
with Pythia 6 for a jet energy $\sim110$ GeV and display the role of hadronization between a 
fictitious partonic shower and a hadronic shower. For the hadronic shower the study includes all
particles in an event.
The shift due to the role of hadronization is very small and can be cross-checked to be the same as
the shift displayed for the energy collimation at $E_{\rm rec}\sim110$ GeV in Fig.~\ref{fig:partonvshadron}. 
However, for energy scales slower than $400$ GeV, the partonic jet shape obtained 
from Pythia 6 is slightly closer to the theoretical calculations.
From the comparison displayed in Fig.~\ref{fig:partonvshadron} 
we can conclude that for high-energy jets the shift between both 
curves is very small and vanishes asymptotically. This is another part of the reason why, 
Pythia 6 and YaJEM+BW with account of hadronization provide a more accurate agreement with the data.
\section{Summary}
In this paper, we studied the energy collimation of gluon and quark 
jets produced in heavy-ion collisions and the jet shape of hadrons 
produced in $pp$ and PbPb collisions at 2.76 TeV. We extracted the LLA and NLLA jet shapes
for quark and gluon jets from the jet energy collimation in the frame of DGLAP
evolution at large $x$ including the scaling violation of FFs 
for the first time in this work. More efforts in the 
numerical framework are however required in order to improve our results and provide
a more accurate description which may improve the shape and normalization
of both observables, as explained in the main body of this paper.

The NLLA energy collimation 
seems to capture a more complete analytical description of this observable than the 
LLA energy collimation obtained in Ref.~\cite{PerezRamos:2012ci}, particularly in gluon jets, 
but a disagreement with YaJEM+BW and the data still persists, which is more pronounced in more biased quark 
showers with smaller jet resolutions, i.e. $R=0.1$. The difference between 
gluon and quark jets for this observable is qualitatively well described by both medium-modified NLLA 
and YaJEM+BW descriptions, i.e. both provide stronger energy collimation in quark jets than in gluon jets
and the NLLA description improves the normalization for partons carrying the intermediate energy fractions $x\sim0.5$. 
Though this quantity cannot be directly measured for each type of jet separately, their combination 
would lead to the quantification of the jet broadening at high-energy heavy-ion 
experiments, i.e. the NLLA formula in the vacuum ($N_s=1$) and Pythia 6 predict a much faster increase of the 
energy collimation than the medium-modified NLLA ($N_s=1.4$) and YaJEM+BW calculations as the energy scale 
increases. 


We extracted the jet shape from the analysis performed for the energy collimation and 
compared the NLLA, LLA, Pythia 6 and YaJEM+BW calculations with the CMS pp and PbPb data 
at 2.76 TeV. The final biased and unbiased comparison for this observable clearly shows 
the importance of taking all jet-finding conditions into account in order to get as 
accurate results as possible in the comparison of Monte Carlo event generators 
and theoretical predictions with the data. 

The NLLA and LLA predictions qualitatively describe the jet shapes but 
fail to reproduce the right normalization of this observable. The reasons for this 
disagreement are the same as those previously presented for the energy collimation
in the last paragraph of Sec. \ref{subsec:quarkcollim1}. 
The biased jet shape provided by Pythia 6 is in very good agreement with $pp$ CMS data and the medium-modified 
biased jet shape from YaJEM+BW qualitatively describes the sub-jet broadening shown by PbPb CMS data
for larger values of $r$, although it is much weaker in CMS data than in the YaJEM+BW result.
Gluon jets produce a wider shower broadening than quark jets but they get even more suppressed 
by biases than quark jets, which clearly dominate the data for $E_{\rm rec,jet}\geq100$ GeV. 
This new example proves that biases appear to strongly suppress the relevant physics of
jet quenching we want to understand and hence, information is lost 
concerning the early stage of jet evolution and its interaction with 
the medium in the study of this observable; indeed, the trigger bias suppresses
the range of possible medium modifications brought by the 
medium-induced soft gluon radiation \cite{Renk:2012ve}. 

Of course our results for the jet shape and comparison with the data 
reflect the characteristics of the BW prescription and hence 
should be compared and improved with calculations using other models or 
more conveniently the ongoing calculations of Refs.~\cite{Blaizot:2013vha,Mehtar-Tani:2014yea} 
(for an interesting review see also Ref.~\cite{CasalderreySolana:2012ef}); 
a comparison with YaJEM-DE \cite{Renk:2011qf} may for instance be desirable.

\section*{Acknowledgements}

R. P.-R. is grateful to Beomsu Chang, DongJo Kim and Norbert Novitzky for their 
expert support on the Monte Carlo analysis and coding techniques, and to Wolfgang Ochs for useful 
discussions and comments on the manuscript. We acknowledge support from the Academy 
researcher program of the Academy 
of Finland, Project No. 130472.
\bibliographystyle{unsrt}
\bibliography{mybib}

\end{document}